\documentclass[groupedaddress,
twocolumn,
showkeys, preprintnumbers,
amsmath,amssymb, aps, pre,
draft,
floatfix,]{revtex4-1}

\usepackage{dynlearn}
\usepackage{subfig}
\usepackage{standalone}
\usepackage{amsthm}
\usepackage{mathrsfs}
\usepackage{graphicx}\usepackage{dcolumn}\usepackage{bm}

\theoremstyle{definition}

\renewcommand{\H}{\operatorname{H}}

\newcommand{\One}{ {\mathbf{1} } }
\newcommand{\MSym}{\MeasSymbol}
\newcommand{\msym}{\meassymbol}

\newcommand{\MxSt}{\AlternateState}
\newcommand{\MxSSet}{\AlternateStateSet}
\newcommand{\MxSMeasure}{\mu}

\newcommand{\mxst}{\eta}
\newcommand{\mxstalt}{\zeta}

\newcommand{\simplex}{\Delta}

\newcommand{\BlackwellMeasure}{\mu_B}

\newcommand{\di}{d_1}
\newcommand{\dsc}{d_\mu}

\newcommand{\LCE}{\lambda}
\newcommand{\LCESpectrum}{\Gamma}
\newcommand{\dLCE}{d_\LCESpectrum}

\newcommand{\IFSLongName}{driven iterated function system}
\newcommand{\IFSacronym}{DIFS}
\newcommand{\IFSacronyms}{DIFSs}
\newcommand{\ambiguityrate}{ambiguity rate}
\newcommand{\ha}{h_a}

\begin{document}

\title{Ambiguity Rate of Hidden Markov Processes}

\author{Alexandra M. Jurgens}
\email{amjurgens@ucdavis.edu}

\author{James P. Crutchfield}
\email{chaos@ucdavis.edu}

\affiliation{Complexity Sciences Center, Physics Department\\
University of California at Davis\\
Davis, California 95616}	

\date{\today}
\bibliographystyle{unsrt}

\begin{abstract}
The \eM is a stochastic process' optimal model---maximally predictive and
minimal in size. It often happens that to optimally predict even
simply-defined processes, probabilistic models---including the \eM---must
employ an uncountably-infinite set of features. To constructively work with
these infinite sets we map the \eM to a place-dependent iterated function
system (IFS)---a stochastic dynamical system. We then introduce the ambiguity
rate that, in conjunction with a process' Shannon entropy rate, determines the
rate at which this set of predictive features must grow to maintain maximal
predictive power. We demonstrate, as a ancillary technical result which stands
on its own, that the ambiguity rate is the (until now missing) correction to
the Lyapunov dimension of an IFS's attractor. For a broad class of complex
processes and for the first time, this then allows calculating their
statistical complexity dimension---the information dimension of the minimal set
of predictive features. 
\end{abstract}


\preprint{\arxiv{2008.12886}}

\maketitle

\section{Introduction}
\label{sec:introduction}

An abiding challenge to scientific inquiry is the nature of \emph{complex
systems}---those that create intricate and delicate patterns through their
internal interplay of stochasticity and determinism. These systems are often
identified by the presence intrinsic instabilities, collectively-interacting
subsystems, and visually-striking emergent structures.

Integrating Turing's computation theory \cite{Turi37a, Shan56c, Mins67},
Shannon's information theory \cite{Shan48a}, and Kolmogorov's dynamical systems
theory \cite{Kolm56b, Kolm65, Kolm83, Kolm59, Sina59}, \emph{computational
mechanics} \cite{Crut12a} introduced a suite of tools to analyze complex systems
in terms of their informational architecture. The \emph{\eM}---its most basic
statistic---is a system's maximally predictive, minimal, and unique model. It
captures a system's generation, storage, and transmission of information.
Quantitatively, the information stored in the \eM's \emph{causal states}---the
minimal set of maximally-predictive features---is a process' \emph{statistical
complexity} $\Cmu$, a measure of the memory resources a system employs to
generate its behavior and organization.

\begin{figure}
\centering
(a)
\includegraphics[width=.18\textwidth]{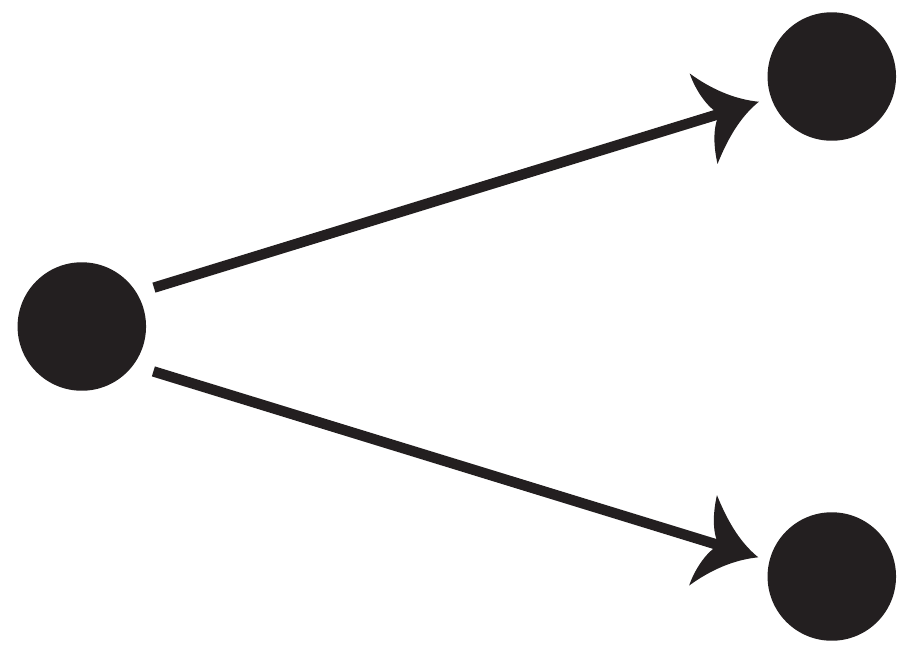}
\hspace{0.25in}
(b)
\includegraphics[width=.18\textwidth]{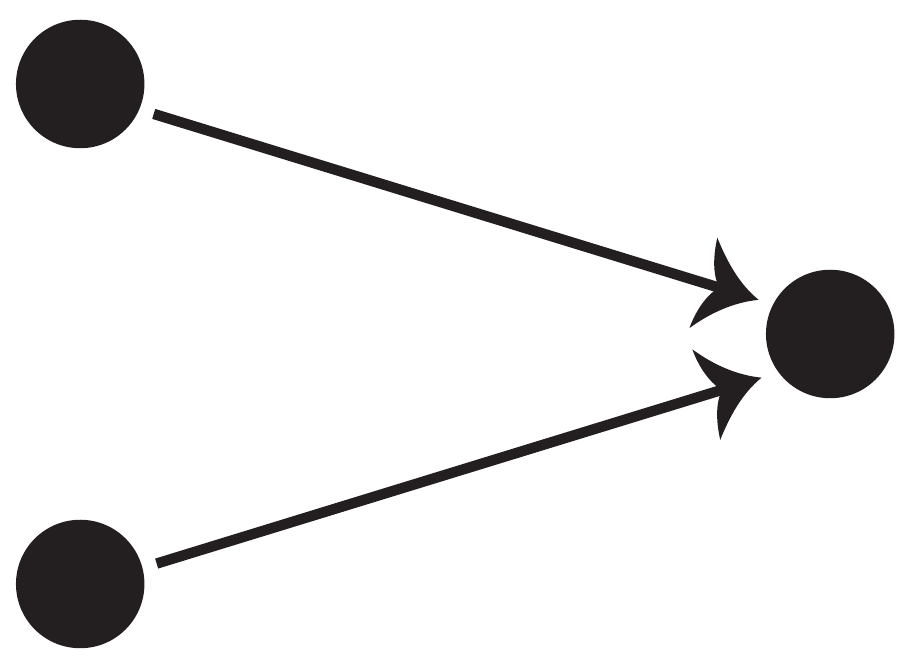}
\caption[text]{(a) Equivocation: Same input sequence leads to different outputs.
	(b) Ambiguity: Two different inputs lead to same output. The strategy
	underlying Shannon's proof of his second coding theorem is to find channel
	inputs that are least ambiguous given the channel's distortion properties,
	which include equivocation.
	}
\label{Fig:AmbiEquivo}
\end{figure}

An optimally predictive model can be imagined as a minimally noisy channel
communicating the system's future into the past. The student of information
theory will recall that Shannon, in his analysis of information transmission
through channels, introduced two mechanisms: \emph{equivocation}, in which the
same input may lead to distinct outputs, and \emph{ambiguity}, in which two
different inputs may lead to the same output; see Fig. \ref{Fig:AmbiEquivo}.
When our channel is taken to be the \eM, the equivocation rate of the channel is
the \emph{entropy rate} $\hmu$ of the underlying system---the rate at which the
system generates future information. This is guaranteed by the predictive
optimality of the \eM---the only noise in the channel is due to the intrinsic
randomness of our complex system.

In the following, we introduce the parallel quantity, the \emph{ambiguity rate}
$\ha$. The ambiguity rate tracks the rate at which the system discards past
information by introducing uncertainty over the infinite past. Explicitly, if a
process can be optimally modeled with a finite set of predictive features, its
\eM must forget information at the same rate at which the system generates it:
$\hmu = \ha$. However, this is atypical, as our predecessor works demonstrated
\cite{Jurg20b, Jurg20c}. In point of fact, for many complex systems, the
predictive-feature set is uncountably infinite and the statistical complexity
$\Cmu$ diverges, requiring the development of new tools to characterize the
complexity of these systems.

The recent work introduced a suite of tools to capture this state of affairs
for a broad class of stochastic processes---those used not only in the study of
complex systems \cite{Crut12a}, but also in coding theory \cite{Marc11a},
stochastic processes \cite{Ephr02a}, stochastic thermodynamics \cite{Bech15a},
speech recognition \cite{Rabi86a}, computational biology \cite{Birney01,
Eddy04}, epidemiology \cite{Breto2009}, and finance \cite{Ryden98}.

The key realization was identifying a process' \eM as the attractor of a hidden
Markov-\emph{Driven Iterated Function System} (\IFSacronym) \cite{Jurg20b}.
First, we showed that this gave efficient and accurate calculation of a
process' Shannon entropy rate $\hmu$. Second, we introduced a new measure of
structural complexity---the \emph{statistical complexity dimension}
$\dsc$---that tracks $\Cmu$'s divergence and gives the information dimension of
the distribution of predictive features \cite{Jurg20c}.

Previously, accurate calculation of $\dsc$ was contingent on the \IFSacronym\
meeting restrictive technical conditions. Introducing \ambiguityrate\ $\ha$
reframes these constraints information-theoretically, effectively lifting them.
The result is a new method to accurately calculate $\dsc$ for a broad class of
complex processes. More abstractly, we propose $\ha$ as a new intrinsic
complexity measure of a stochastic process---the \emph{growth rate} of the
information stored in a process' optimally predictive features. When $\hmu =
\ha$, this growth rate vanishes and the associated \eM's internal causal-state
process is stationary. However, when $\hmu > \ha$, the latter process is
nonstationary and any optimal predictor must accumulate new information over
time to sustain accurate predictions.

Our development below introduces and motivates the \ambiguityrate\ $\ha$.
Sections \ref{sec:Processes} and \ref{sec:InfoTheory} review stochastic
processes and information theory, respectively, and may be skipped by the
familiar reader. \Cref{sec:IteratedFunctionSystems} introduces hidden
Markov-driven iterated function systems.
\Cref{sec:StatisticalComplexityDimension} then discusses the statistical
complexity dimension $\dsc$ and the \emph{overlap problem}---a long-standing
issue in the dimension theory of iterated function systems.
\Cref{sec:BackEntropy} introduces $\ha$ from an information-theoretic
perspective, motivating it as a solution to and a measure of the overlap
problem. Various interpretations are explored, including an historical note on
Shannon's original \emph{dimension rate} from 1948. Finally, to illustrate our
algorithm's effectiveness and the challenges for very complex processes,
\cref{sec:Examples} works through multiple example processes, including
those generated by stationary and nonstationary \eMs.

\section{Processes}
\label{sec:Processes}

A \emph{stochastic process} $\Process$ is a probability measure over a
bi-infinite chain $\ldots \, \MSym_{t-2} \, \MSym_{t-1} \, \MSym_{t} \,
\MSym_{t+1} \, \MSym_{t+2} \ldots$ of random variables, each $\MSym_t$ denoted
by a capital letter. A particular \emph{realization} $\ldots \, \msym_{t-2} \,
\msym_{t-1} \, \msym_{t} \, \msym_{t+1} \, \msym_{t+2} \ldots$ is denoted via
lowercase. We assume values $\msym_{t}$ belong to a discrete alphabet
$\MeasAlphabet$. We work with blocks $\MS{t}{t^\prime}$, where the first index
is inclusive and the second exclusive: $\MS{t}{t^\prime} = \MSym_{t} \ldots
\MSym_{t^\prime-1}$. $\Process$'s measure is defined via the collection of
distributions over blocks: $\{ \Pr(\MS{t}{t^\prime}): t < t^\prime, t,t^\prime
\in \mathbb{Z} \}$.

To simplify, we restrict to stationary, ergodic processes: those for which
$\Prob(\MS{t}{t+\ell}) = \Prob(\MS{0}{\ell})$ for all $t \in \mathbb{Z}$, $\ell
\in \mathbb{Z}^+$, and for which individual realizations obey all of those
statistics. In such cases, we only need to consider a process's length-$\ell$
\emph{word distributions} $\Prob(\MS{0}{\ell})$.

A \emph{Markov process} is one for which $\Pr(\MSym_t|\MS{-\infty}{t}) =
\Pr(\MSym_t|\MSym_{t-1})$. A \emph{hidden Markov process} is the output of a
memoryless channel \cite{Cove06a} whose input is a Markov process
\cite{Ephr02a}.

\section{Information Theory}
\label{sec:InfoTheory}

Beyond its vast technological applications to communication systems
\cite{Cove06a}, Shannon's information theory \cite{Shan48a} is a widely-used
foundational framework that provides tools to describe how stochastic processes
generate, store, and transmit information. In particular, we use information
theory to study complex systems as it makes minimal assumptions as to the
nature of correlations between random variables and handles multi-way,
nonlinear correlations that are common in complex processes. Here, we now
briefly recall several concepts needed in the following.

Information theory's most basic measure is the Shannon \emph{entropy}.
Intuitively, it is the amount of information that one gains when observing a
sample of a random variable. Equivalently modulo sign, it is also the amount
of uncertainty one faces when predicting the sample.  The entropy $H[\MSym]$ of
the random variable $\MSym$ is: 
\begin{align}
H[\MSym] =
  - \sum_{\msym \in \MeasAlphabet} \Pr(\MSym = \msym) \log_2 \Pr(\MSym = \msym)
  ~.
\label{eq:ShannonEntropy}
\end{align}
We can probe the relationship between two jointly-distributed random variables,
say, $\MSym$ and $Y$. There is the \emph{joint entropy} $H[\MSym, Y]$, of the
same functional form but applied to the joint distribution $\Pr(\MSym,Y)$. And,
there is \emph{conditional entropy} that gives the amount of information
learned from observation of one random variable given another:
\begin{align}
  H[\MSym | Y ] = H[\MSym , Y] - H[Y]  ~. 
  \label{eq:ConditionaLEntropy}
\end{align}

Conditional entropy can be generalized to describe processes in terms of the
\emph{intrinsic} randomness---the amount of information one learns upon
observing the next emitted symbol $X_0$, given complete knowledge of the
infinite past. This is the Shannon \emph{entropy rate}:
\begin{align}
  \hmu = \lim_{\ell \to \infty} H[\MSym_0 | \MSym_{-\ell : 0 }]
~,
\label{eq:ShanmonEntropyRate}
\end{align}
the irreducible amount of information gained in each time step. 

The fundamental measure of correlation between random variables is the
\emph{mutual information}. It can be written in terms of Shannon entropies:
\begin{align}
  I[\MSym ; Y] = H[\MSym, Y] - H[\MSym | Y] - H[Y | \MSym] ~.
  \label{eq:MutualInformation}
\end{align}
As should be clear by inspection, the mutual information between two variables
is symmetric. When $\MSym$ and $Y$ are independent, the mutual information
between them vanishes. As with entropy, we may condition the mutual information
on another random variable, giving the \emph{conditional mutual information}:
\begin{align}
  I[\MSym ; Y | Z] = H[\MSym | Z ] + H[Y | Z] - H[\MSym, Y | Z] ~.
  \label{eq:ConditionalMutualInformation}
\end{align}
The conditional mutual information is the amount of information shared by
$\MSym$ and $Y$, given we know a third, $Z$. Note that $\MSym$ and $Y$ can
share mutual information, but be conditionally independent. Moreover,
conditioning on a third variable $Z$ can either increase or decrease mutual
information \cite{Cove06a}. That is, two variables can appear more or less
dependent, given additional data.

\section{\IFSLongName}
\label{sec:IteratedFunctionSystems}

Our main objects of study are hidden Markov processes. The following introduces
\IFSLongName\ as a class of predictive models for them. A given hidden Markov
process can have many alternative models, each is referred to as a
\emph{presentation}. Driven iterated functions systems are one class of
presentations.

\newcommand{\HMMStates}{\mathcal{V}}

\begin{Def}
\label{Def:\IFSacronym}
An $N$-dimensional hidden Markov-\emph{\IFSLongName}
(\IFSacronym) $\left( \MeasAlphabet, \HMMStates, \MxSSet, \{T^{(\msym)}\}, \{p^{(\msym)}\},
\{f^{(\msym)}\}: \msym \in \MeasAlphabet \right)$ consists of:
\begin{enumerate}
\setlength{\topsep}{0mm}
\setlength{\itemsep}{0mm}
\item a finite alphabet $\MeasAlphabet$ of $k$ \emph{symbols}
	$x \in \MeasAlphabet$,
\item a set $\HMMStates$ of $N$ \emph{presentation states},
\item a set of \emph{states} $\MxSSet \subset \simplex^{(N-1)}$, over
	$N$-dimensional presentation-state distributions $\mxst \in \MxSSet$, 
\item a finite set of $N$ by $N$ \emph{symbol-labeled substochastic matrices}
	$T^{(\msym)}$, $\msym \in \MeasAlphabet$,
\item a set of $k$ \emph{symbol-labeled probability functions} 
  $p^{(\msym)} = \langle \mxst | T^{(\msym)} \One \rangle $, and
\item a set of $k$ \emph{symbol-labeled mapping functions} 
  $f^{(\msym)} = \langle \mxst | T^{(\msym)} \One \rangle
  / p^{(\msym)}(\mxst)$.
\end{enumerate}
\end{Def}

The \emph{(N-1)-simplex} $\simplex^{N-1}$ is the set of presentation-state
probability distributions such that:
\begin{align*}
 \{ \eta \in \mathbb{R}^{N} : \langle\eta\ket{\One} = 1, \langle\eta\ket{\delta_i} \geq 0,
  i=1, \dots , N \}
  ~,
\end{align*}
where $\bra{\delta_i} = \begin{pmatrix} 0 & 0 & \dots & 1 & \dots & 0
\end{pmatrix}$---that is, $\bra{\delta_i}_i = 1$, otherwise $0$---and
$\ket{\One} = \begin{pmatrix} 1 & 1 & \dots & 1 \end{pmatrix}$. We use this
notation for components of the presentation-state vector $\mxst$ to avoid
confusion with temporal indexing.

\begin{figure}[ht]
\centering
\includegraphics[width=.45\textwidth]{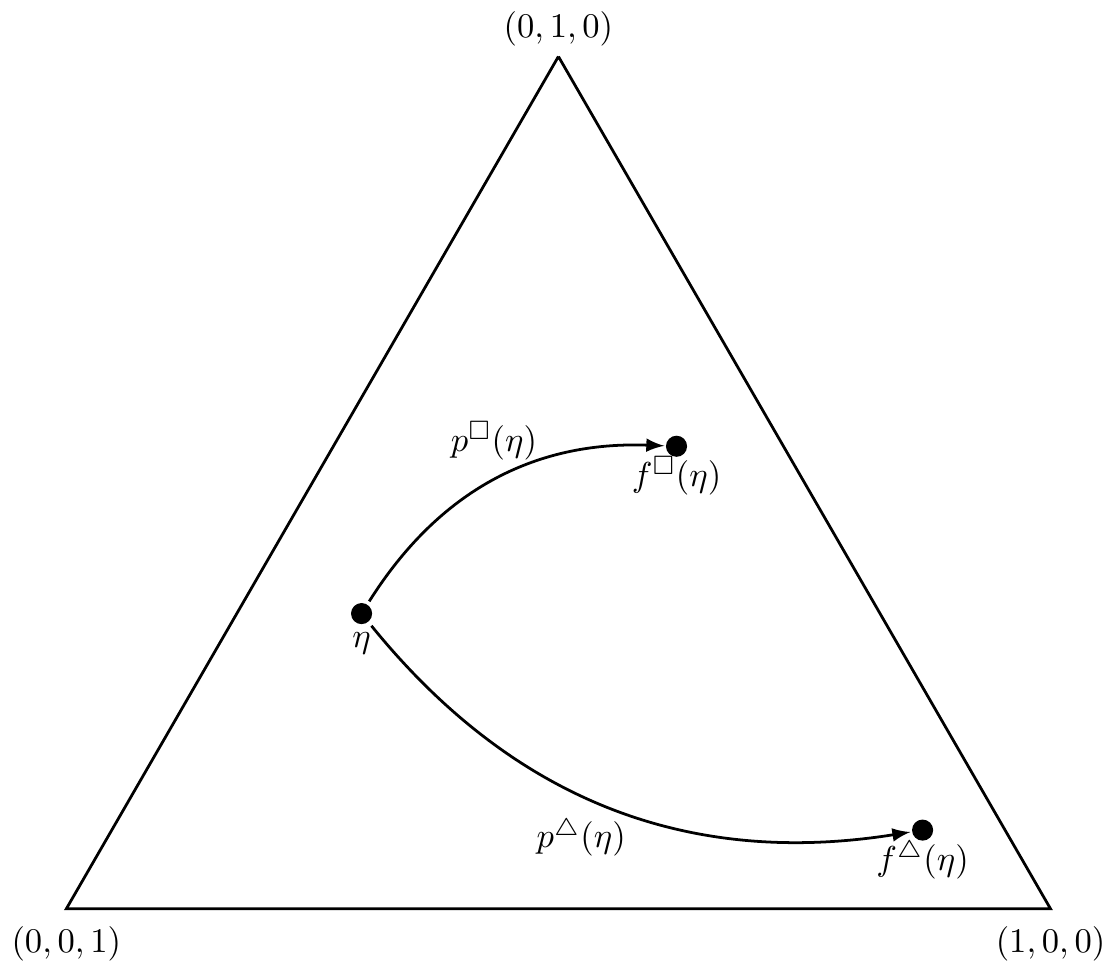}
\caption[text]{How a hidden Markov-driven iterated function system
  	(\IFSacronym) generates a hidden Markov process: An initial state
	$\mxst$---a distribution over three states: $(0,0,1)$, $(0,1,0)$, and
	$(1,0,0)$---in the $2$-simplex is associated with a transition probability
	distribution $p^\msym (\eta)$ over the alphabet $\msym \in \MeasAlphabet =
	\{ \square, \triangle \}$. If the emitted symbol selected from this
	distribution is $\square$, the next state is generated according to the
	associated mapping function $f^{(\square)}(\mxst)$ and the probability
	distribution is updated accordingly. The same steps are followed if the
	symbol is $\triangle$ using $f^{(\triangle)}(\mxst)$, resulting in an
	emitted process $\Process$ over symbols $\MeasAlphabet$.
	}
\label{Fig:IFScartoon}
\end{figure}

The set of substochastic matrices must sum to the nonnegative, row-stochastic
matrix $T = \sum_{\msym \in \MeasAlphabet} T^{(\msym)}$---the transition matrix
for the presentation-state Markov chain. This ensures that $\sum_{\msym \in
\MeasAlphabet} p^{(\msym)} (\mxst) = 1$ for all $\mxst \in \simplex^{(N-1)}$. 

\Cref{Fig:IFScartoon} shows how a \IFSacronym\ generates a hidden Markov
process: Given an initial state $\mxst_0 \in \simplex^{N-1}$, the probability
distribution $\{ p^{(\msym)}(\mxst_0) : \msym = 1, \dots, k \}$ is sampled.
According to the realization $\msym_0$, apply the mapping function to map
$\mxst_0$ to the next state $\mxst_1 = f^{(\msym_0)}(\mxst_0)$. According to
the new probability distribution defined by $\mxst_1$, draw $\msym_1$ and
repeat. This action generates our emitted process $\Process$: $\msym_0,
\msym_1, \msym_2, \ldots$.

\begin{figure*}
  \centering
  \includegraphics[width=.95\textwidth]{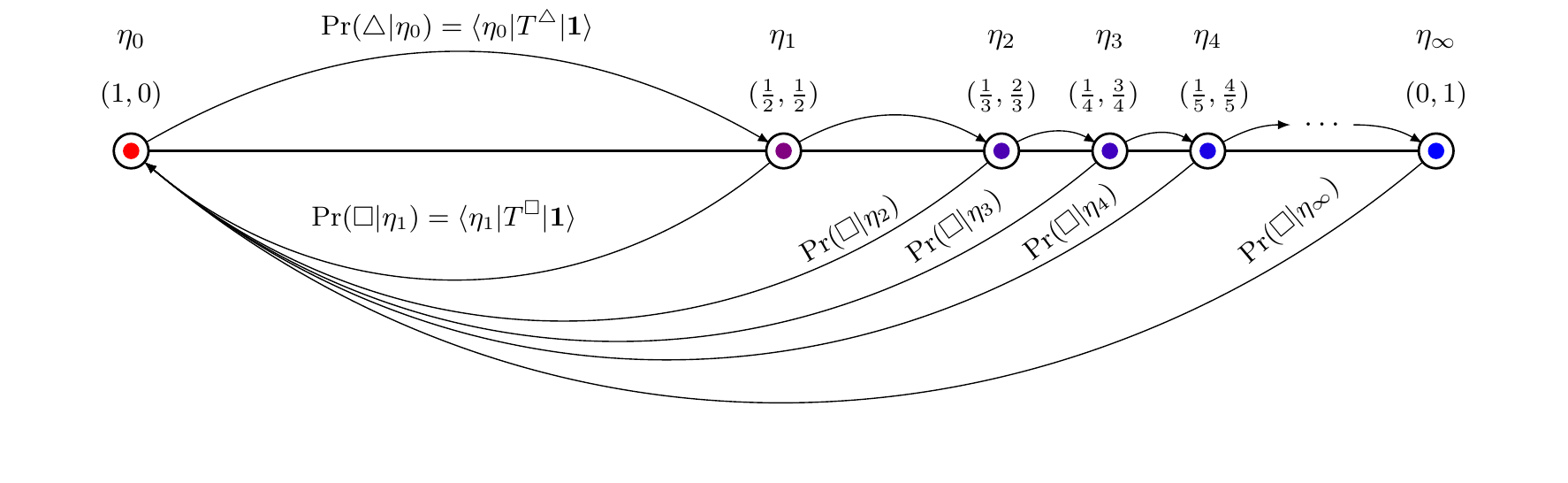}
  \caption[text]{The states and transitions of a hidden Markov-driven iterated
  function system (\IFSacronym) discussed in \cref{sec:exampleSNS} embedded in
  the 1-simplex. In this case, the set of states $\MxSSet$ is countable, which
  each subsequent application of $f^{(\triangle)}$ bringing $\mxst$ nearer to
  $(0,1)$, which is reached only after observing infinitely many $\triangle$s.
  The countable nature of the state set arises from the the fact that one of the
  mapping functions is a constant: $f^{(\square)} = (1,0)$.}
  \label{Fig:SNSsimplexEmbedding}
\end{figure*}

\begin{figure*}
  \centering
\subfloat[Set of states generated by the ``delta'' \IFSacronym. 
  \label{fig:simplex_example_delta}]{
      \includegraphics[width=0.32\textwidth]{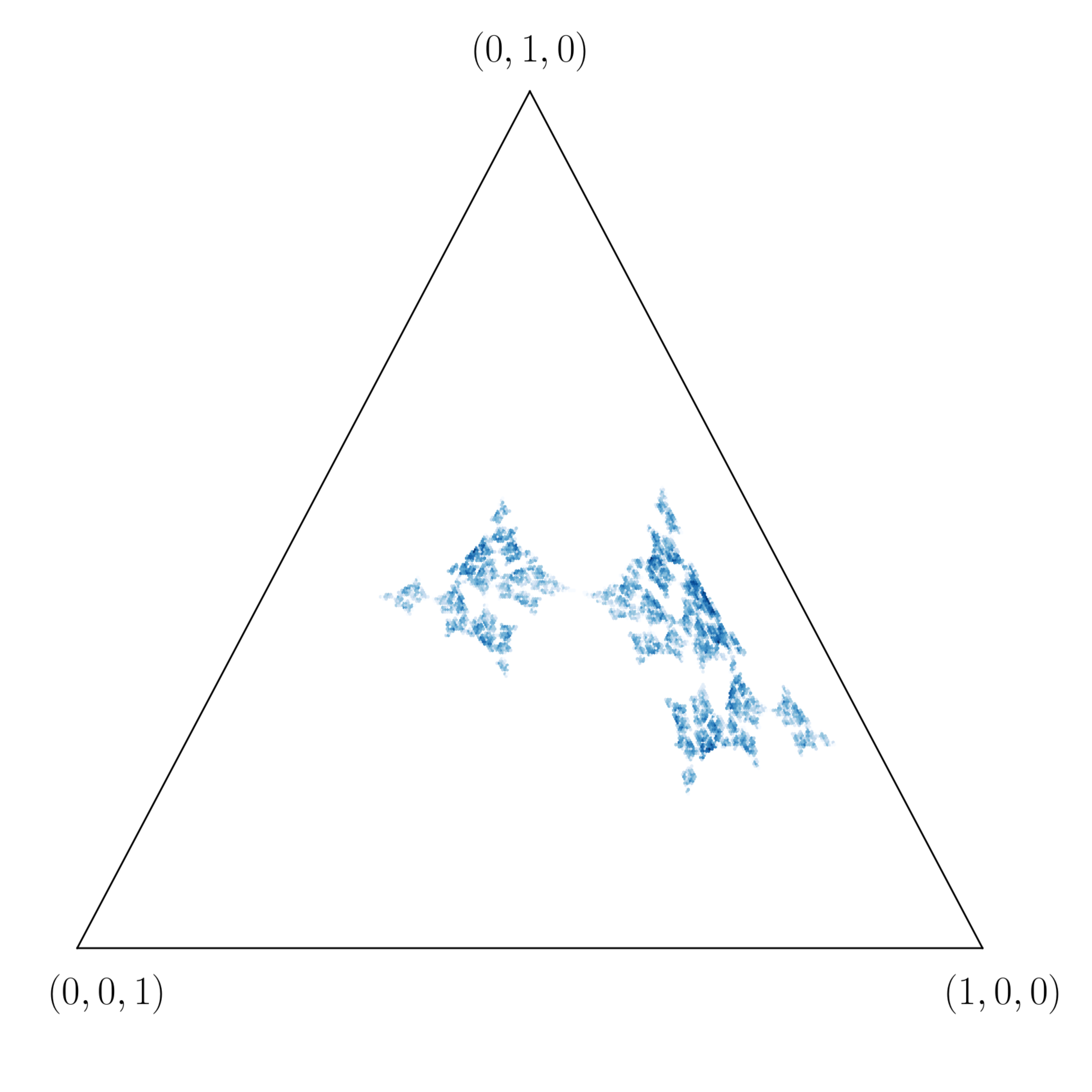}
  } 
\subfloat[Set of states generated by the ``Nemo'' \IFSacronym. 
  \label{fig:simplex_example_nemo}]{
      \includegraphics[width=0.32\textwidth]{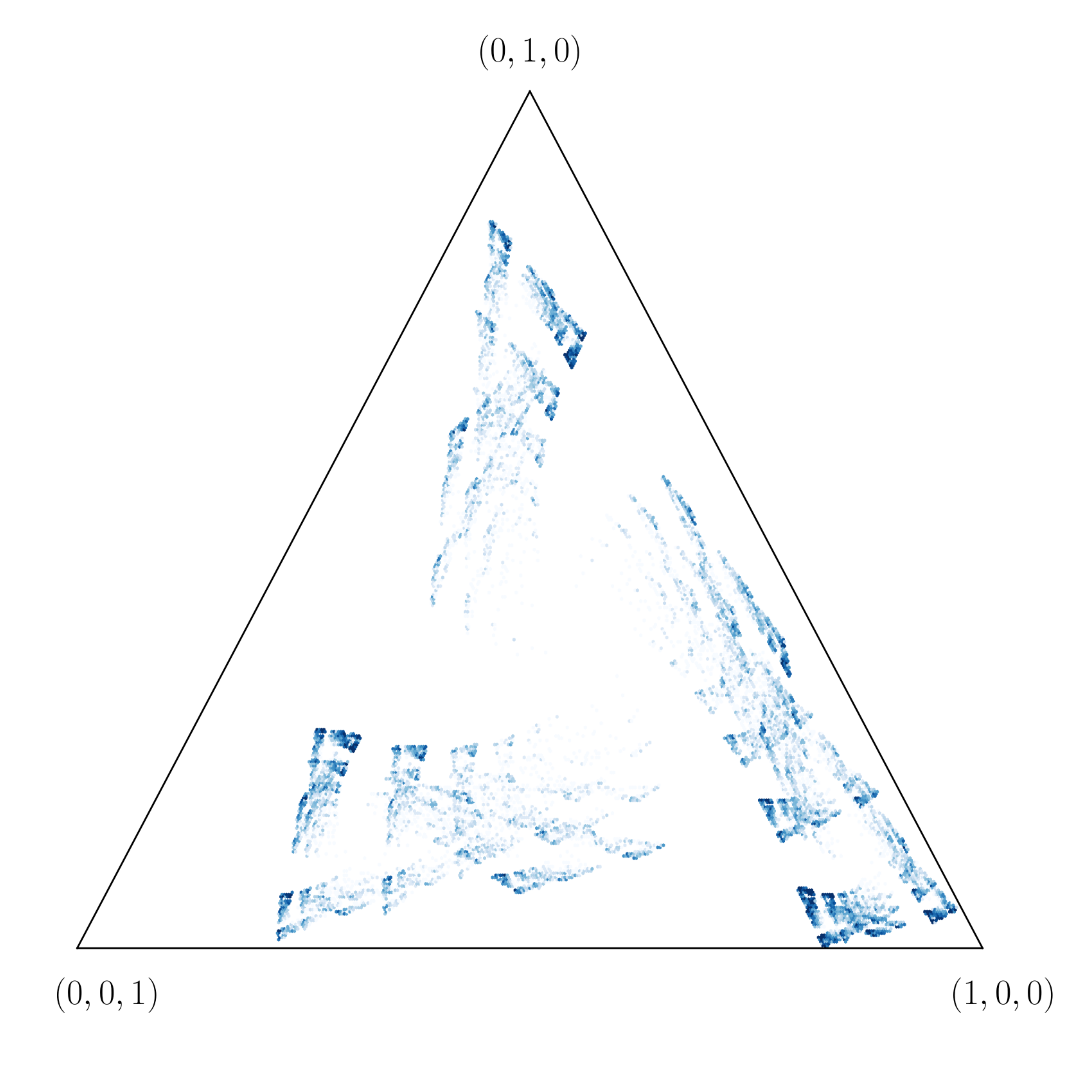}
  }
\subfloat[Set of states generated by the ``gamma'' \IFSacronym. 
  \label{fig:simplex_example_gamma}]{
      \includegraphics[width=0.32\textwidth]{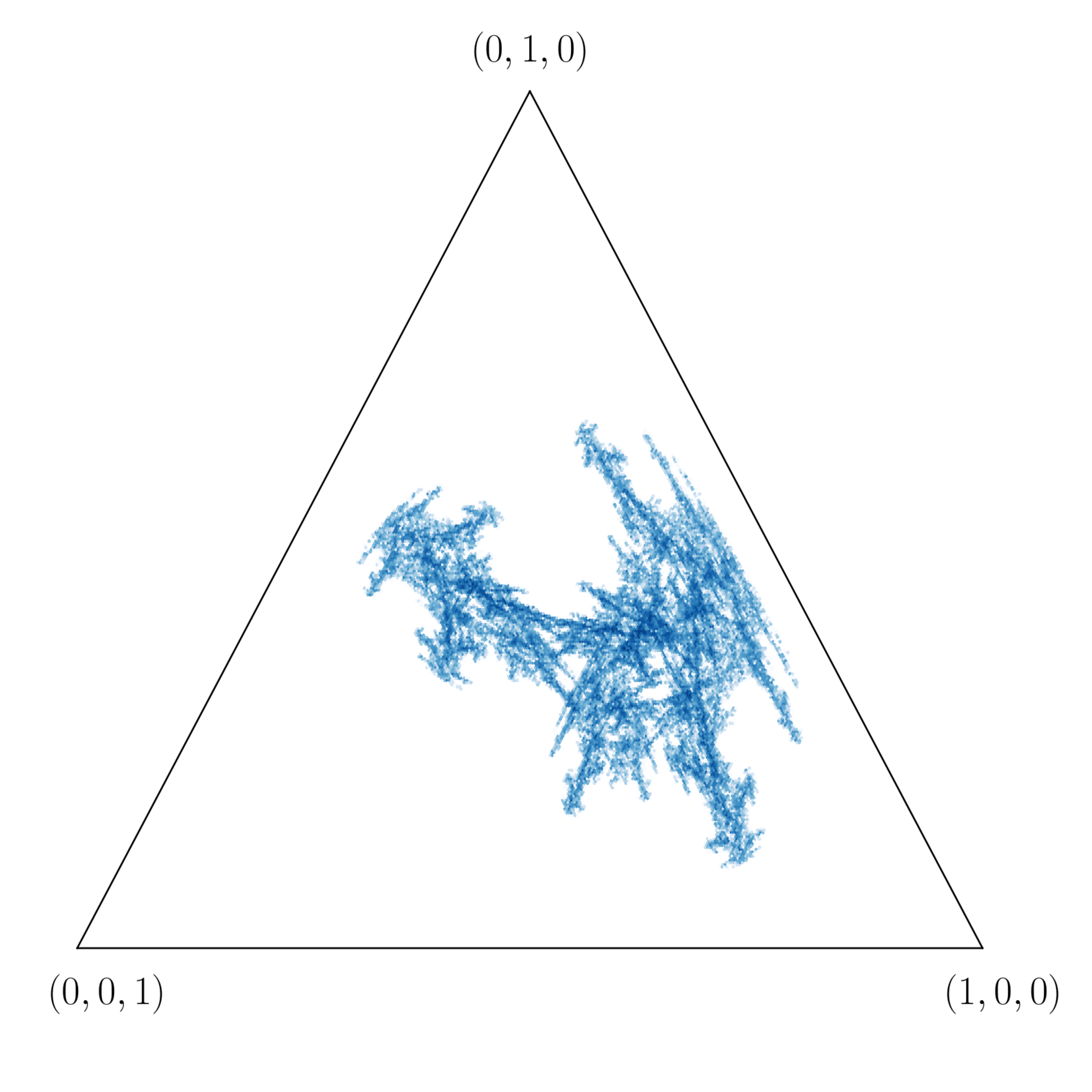}
  } 
\caption{Hidden Markov \IFSLongName\ (\IFSacronym) may generate state sets with
	a wide variety of structures, many fractal in nature. Each subplot displays
	$10^5$ states of a different \IFSacronym. The \IFSacronym s themselves are
	specified in \cref{app:nonunifilarHMCs}.}
\label{fig:simplex_examples}
\end{figure*}

This describes the random dynamical system---the DIFS---that generates the
hidden state sequence $\mxst_0, \mxst_1, \mxst_2, \ldots$. As we previously
showed, the attractor of this dynamical system is the invariant set of states
$\MxSSet$ and their evolution is ergodic \cite{Jurg20b,Elton1987}.
Additionally, the attractor has a unique, attracting, invariant measure known
as the \emph{Blackwell measure} $\BlackwellMeasure(\MxSSet)$ \cite{Blac57b}.
Although $\MxSSet$ may be countable, as for the \IFSacronym\ depicted in
\cref{Fig:SNSsimplexEmbedding}, in general, $\MxSSet$ will be uncountably
infinite and fractal in nature, as in the examples in
\cref{fig:simplex_examples}.

\IFSacronym\ states are \emph{predictive} in the sense that they are functions
of the prior sequences of observables (pasts) and lead to the correct future
distribution conditioned on the pasts. Consider an infinitely-long past that,
in the present, has induced some state $\mxst$. It is not guaranteed that this
infinitely-long past induce a unique state, but it is the case that any state
induced by this past must have the same conditional future distribution.
Indeed, for task of prediction, knowing the previous state is as good as
knowing the infinite past: $\Pr(\MSym_{0: \ell} | \MxSt_0 = \mxst) = \Pr(
\MSym_{0: \ell} | \MSym_{-\infty : 0})$ for all $\ell \in \mathbf{N}^+$.

Therefore, the \IFSacronym\ is a predictive model of the given process
$\Process$. Contrast this to it being merely a \emph{generative model}
that produces all and only the sequences in a process, but whose states need
not be predictive. Note that predictive models are generative.

Borrowing from the language of automata theory, we refer to the set of states
$\MxSSet$ plus its transition dynamic---$\Pr(\msym_t| \mxst_t)$ and
$\Pr(\mxst_{t+1} | \mxst_t, \msym_t)$---as a \emph{state machine} or, simply
\emph{machine} that optimally predicts $\Process$. When we force each
infinitely long past to induce a unique state, we produce a canonical
predictive model that is unique: a process' \emph{\eM} \cite{Crut12a}.

\begin{Def}
An \emph{\eM} is a \IFSacronym\ with \emph{probabilistically distinct states}: For each
pair of distinct states $\mxst, \mxstalt \in \MxSSet$
there exists a finite word $w = \ms{0}{\ell-1}$ such that: 
\begin{align*}
\Prob(\MS{0}{\ell} = w|\MxSt_0 = \mxst)
  \not= \Prob(\MS{0}{\ell} = w|\MxSt_0 = \mxstalt)~.
\end{align*}
\label{def:eM}
\end{Def}

A process' \eM\ is its optimally-predictive, minimal model, in the sense that
the set $\MxSSet$ of predictive states is minimal compared to all its other
predictive models. By capturing a process' structure and not merely being
predictive, an \eM's states are called \emph{causal states}. Unless otherwise
noted, we assume that all \IFSacronym\ discussed here are $\epsilon$-machines.

Calculating the Shannon entropy rate for a process generated by an \IFSacronym\
was the focus of our first analysis of \IFSacronym s \cite{Jurg20b}. Due to
the associated process' ergodicity, shown there, $\hmu$ may be written:
\begin{align}
  \widehat{\hmu}^B = - \lim_{\ell \to \infty} \frac{1}{\ell} 
  \sum_{t = 0}^\ell \sum_{\msym \in \MeasAlphabet} \Pr(\msym |\mxst_\ell) 
  \log_2 \Pr(\msym |\mxst_\ell)
   ~.
 \label{eq:blackwellentropylimit}
 \end{align}
This tracks the uncertainty in the next symbol $\msym$ given our current causal
state $\mxst_t$, averaged over the Blackwell measure; hence the superscript
$B$. It quantifies the intrinsic randomness of the process $\Process$. 

\section{Statistical Complexity Dimension}
\label{sec:StatisticalComplexityDimension}

Images of the self-similar \IFSacronyms\ state sets $\MxSSet$ (as in
\cref{fig:simplex_examples}) are evocative and lead naturally to questions
about how $\MxSSet$'s geometric properties relate to intrinsic properties of
the underlying process $\Process$. To answer this, we say that a process'
memory is the information required to specify its \eM states, i.e., the minimal
amount of information needed to predict $\Process$. This may be measured either
in terms of the cardinality $|\MxSSet|$ of causal states or the amount of
historical Shannon entropy they store---that is, the \emph{statistical
complexity} $\Cmu$.

\begin{Def}
\label{def:Cmu}
A process' \emph{statistical complexity} is the Shannon entropy stored in its
\eM's causal states:
\begin{align}
	\nonumber
\Cmu = & \H[\Pr(\MxSSet)] \\
     = & - \sum_{\mxst \in \MxSSet}
	 p(\mxst) \log_2 p(\mxst)
  ~.
\label{eq:Cmu}
\end{align}
\end{Def}

From the definitions above, a process' \eM\ is its smallest predictive model, in
the sense that both |$\MxSSet|$ and $\Cmu$ are minimized by a process' \eM,
compared to all other predictive models. Due to the \eM's unique minimality, we
identify the \eM's $\Cmu$ as the process' memory.

However, when the set of causal states $\MxSSet$ is infinite, the statistical
complexity may diverge. In this case, $\Cmu$ is no longer an appropriate
complexity measure to distinguish processes. Despite this, a need remains: It
is clear that processes with infinite state sets differ significantly in
internal structure, as shown in \cref{fig:simplex_examples}. In this case, we
turn to the \emph{statistical complexity dimension}, defined as the rate of
divergence of the statistical complexity, to serve as a measure of structural
complexity. This leaves us with an abiding question, though, What does it mean
that a finitely-specified process' state information (memory) diverges?

\subsection{Dimension and Causal State Divergence}
\label{sec:FractalDimension}

A set's \emph{dimension}, construed most broadly, gives the rate at which a
chosen size-metric diverges with the scale at which the set is observed
\cite{Reny59a,Mand82a,Edga90a,Falc90a,Pesi97a}. Fractional dimensions, in
particular, are useful to probe the ``size'' of sets when cardinality alone
is not informative. ``Fractal dimension'', said in isolation, is often taken to
refer to the box-counting or Minkowski-Bouligand dimension. The following,
though, determines the \emph{information dimension}---a dimension that accounts
for the scaling of a measure on a fractional dimension set. In this case, our
measure of interest is the Blackwell measure $\BlackwellMeasure$ over our
causal states $\MxSSet$.

Consider the state set $\MxSSet$ on the $(N-1)$-simplex for an \IFSacronym\ that
generates a process $\Process$. Coarse-grain the $N$-simplex with evenly-spaced
subsimplex cells of side length $\epsilon$. Let $\mathcal{F}(\epsilon)$ be the
set of cells that encompass at least one state. Now, let each cell in
$\mathcal{F}(\epsilon)$ itself be a (coarse-grained) state and approximate the
\eM dynamic by grouping all transitions to and from states encompassed by the
same cell. This results in a finite-state Markov chain that generates an
approximation of the original process $\Process$ and has a stationary
distribution $\mu(\mathcal{F}(\epsilon))$. Then $\BlackwellMeasure(\MxSSet)$'s
\emph{information dimension} is:
\begin{align}
\di (\BlackwellMeasure(\MxSSet) )
  & =  \lim_{\epsilon \to 0} 
  \frac{ H_{\MxSMeasure} [\mathcal{F}(\epsilon)] }{ \log \epsilon }
  ~,
\label{eq:InformationDimension}
\end{align}
where $H_{\MxSMeasure} [\mathcal{F}(\epsilon)] = - \sum_{C_i \in
\mathcal{F}(\epsilon)} \MxSMeasure(C_i) \log \MxSMeasure (C_i)$ is the Shannon
entropy over the set $\mathcal{F}(\epsilon)$ of cells that cover attractor
$\MxSSet$ with respect to $\MxSMeasure$.

Rearranging \cref{eq:InformationDimension} shows that the state entropy of the
finite-state approximation scales logarithmically with $\MxSSet$'s information
dimension with respect to the Blackwell measure:
\begin{align}
    H_\mu [\mathcal{F}] \sim \di (\BlackwellMeasure) \cdot \log \epsilon
	~.
\label{eq:StateEntropyScalingRelationship}
\end{align}
Applied to a process $\Process$'s \eM, $\di$ describes the divergence rate of
statistical complexity $\Cmu$:
\begin{align}
  \Cmu(\epsilon) \sim \dsc \cdot \log \epsilon
  ~.
\label{eq:CmuScalingRelationship}
\end{align}
In this way, we refer to the \eM's information dimension
$\di(\BlackwellMeasure)$ as $\Process$'s \emph{statistical complexity
dimension} $\dsc$.

\subsection{Determining Statistical Complexity Dimension}
\label{sec:Calculatingdsc}

Directly calculating the statistical complexity dimension using
\cref{eq:InformationDimension} is nontrivial, as it often requires estimating a
fractal measure. Fortunately, as two previous works discussed and as we now
show, the intractability can be circumvented by leveraging the process's
associated generating dynamical system---the \IFSacronym---to calculate $\dsc$
\cite{Jurg20b, Jurg20c}.

For a dynamical system, the \emph{spectrum of Lyapunov characteristic
exponents} $\LCESpectrum = \{\LCE_1, \ldots, \LCE_N : \LCE_i \geq \LCE_{i+1}
\}$ \cite{Shim79a,Bene80a} measures expansion and contraction as the average
local growth or decay rate, respectively, of orbit perturbations. The result is
a list of rates that indicate long-term orbit instability ($\LCE_i > 0)$ and
orbit stability ($\LCE_i < 0)$ in complementary directions.

Consider covering an attractor generated by a dynamical system $f$ with
hypercubes of side length $\epsilon$. After applying $f$ to a hypercube $k$
times, the side lengths are approximately $\epsilon e^{\LCE_1 k}, \epsilon
e^{\LCE_2 k}, \dots$, assuming that the hypercube orientation is chosen
appropriately. This property allows combining the $\LCESpectrum$ into a
expression approximating the growth rate of hypercubes needed to cover the
attractor, as $\epsilon \to 0$. In turn, this implies a natural relationship
between the $\LCESpectrum$ and dimensional quantities, such as
\cref{eq:InformationDimension}. In point of fact, the \emph{Lyapunov dimension}
\cite{Kapl79a} has been conjectured to be equivalent to the information
dimension for ``typical systems'' $f$.

Our previous work showed how to calculate $\LCESpectrum$ for \IFSacronym s
\cite{Jurg20c}. However, since \IFSacronym s are \emph{random} dynamical
systems, additional orbit expansion arises from the stochastic selection of the
maps $f^{(\msym)}$. Indeed, for \IFSacronyms, since the maps are contractive
\emph{all} expansion arises from this stochastic choice, which is measured by
the Shannon entropy rate $\hmu$ of the generated process $\Process$. That is to
say, for \IFSacronym s, $\LCE_i < 0$ for all $i$ while $\hmu$ monitors the
expansive exponent.

\begin{figure}
  \centering
  \subfloat[ 
    \label{fig:simplex_example_nonoverlapping}]{
        \includegraphics[width=0.47\columnwidth]{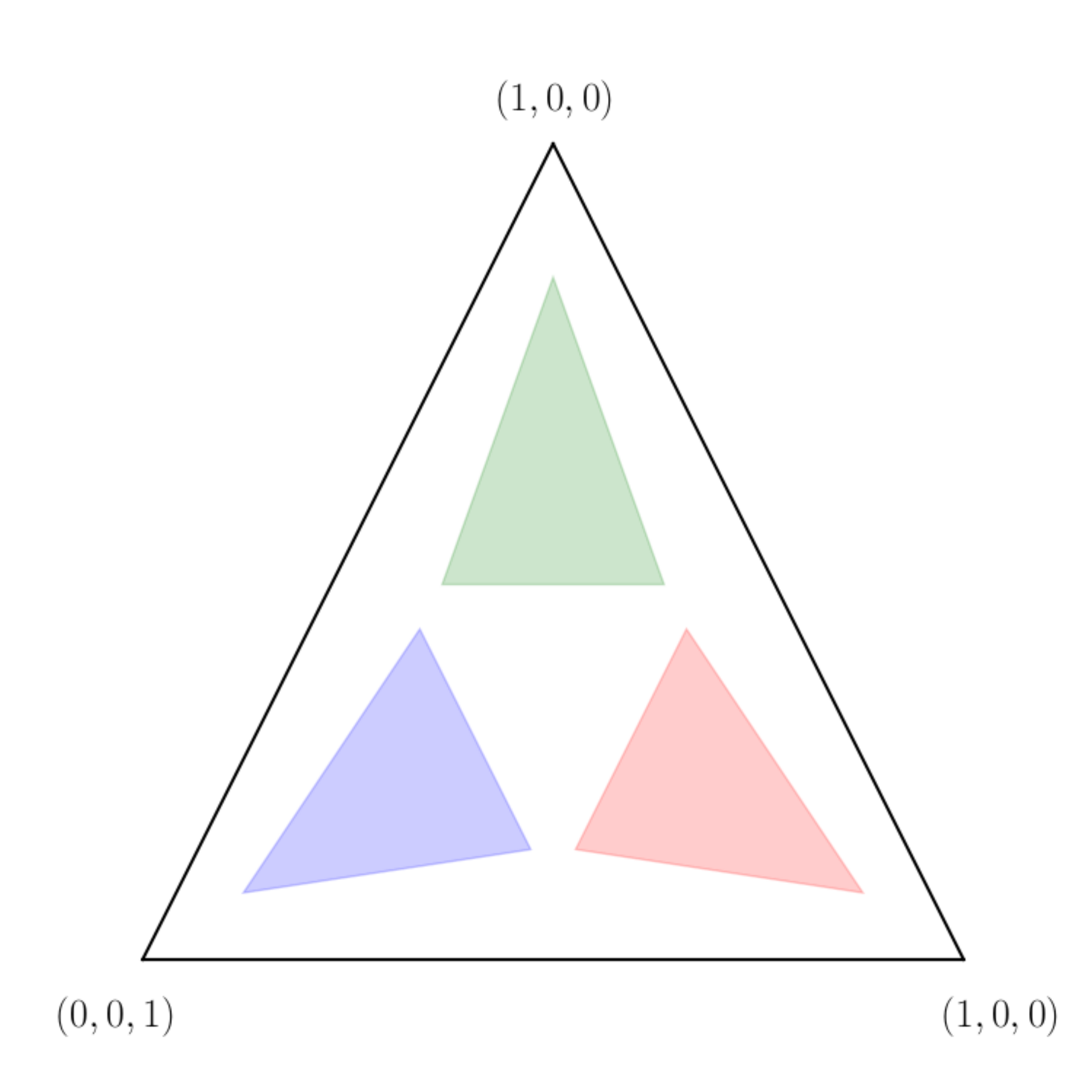}
    }
  \subfloat[
    \label{fig:simplex_example_overlapping}]{
        \includegraphics[width=0.47\columnwidth]{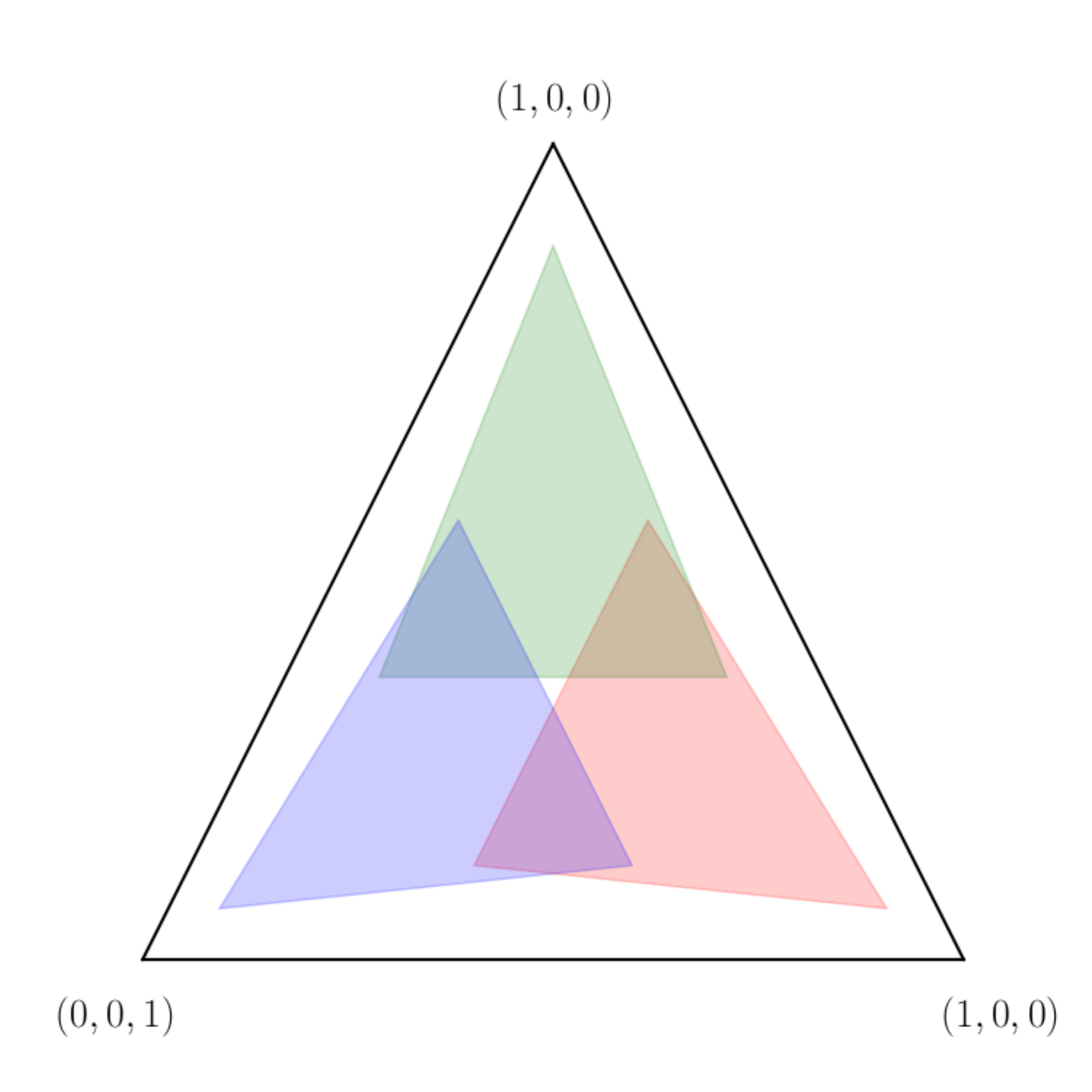}
    } 
  \caption[text]{Overlap problem on the $2$-simplex $\simplex^2$: Two distinct
    \IFSacronym s (given in \cref{app:nonunifilarHMCs}) are considered, each
    with three mapping functions. Images of the mapping functions over the
    entire simplex are depicted as regions in red, blue, and green. (a) Images
    of the mapping functions $f^{(\triangle)}$,$f^{(\square)}$, and
    $f^{(\circ)}$ do not overlap---every possible state has a unique pre-image.
    (b) Images of the mapping functions coincide: there exist $\eta_1, \eta_2
    \in \simplex^2$ such that $f^{'(\triangle)}(\eta_1) =
    f^{'(\square)}(\eta_2) = \eta_3$. This case is an \emph{overlapping
    \IFSacronym}.
    }
  \label{Fig:simplex_overlap}
  \end{figure}
  
With this in mind, we adapt the Lyapunov dimension expression to \IFSacronym
s as follows: 
\begin{align}
\widetilde{\dLCE} = \left\{
  \begin{aligned}
	& k + \frac{\Lambda (k) + \hmu}{|\lambda_{k+1}|} ,
	& -\Lambda (N) > \hmu  \\
	& N ~,
	& - \Lambda (N) \leq \hmu
  \end{aligned}
  \right.
  ~,
\label{eq:IFSLyapunovDimension}
\end{align}
where we introduce the Lyapunov spectrum partial sum $\Lambda (m) = \sum_{i=1}^m
\lambda_i$ and $k = 0, 1, 2, \ldots, N-1$ is the largest index for which $-
\Lambda (k) < \hmu$. Note $\Lambda(m) < 0, m = 1, 2, \ldots, N$ and we take
$\Lambda (0) = 0$. Readers familiar with the Lyapunov dimension should take care
as we have re-indexed from the traditional presentation of $\dLCE$ for
readability. 

Under specific technical conditions, $\dLCE$ is exactly the information
dimension of the \IFSacronym's attractor: $\dLCE = \dsc$ \cite{Baran15b}.
Generally, relaxing the conditions, $\widetilde{\dLCE}$ only upper bounds the
statistical complexity dimension:
\begin{align}
  \widetilde{\dLCE} \geq \dsc
  ~. 
\label{eq:DSCbound}
\end{align}
The extent to which the bound is not saturated is in large part determined by
the \emph{open set condition}, which we now discuss. We, then, turn to solve the
associated ``overlap problem''. This leads to an exact expression for
\IFSacronym\ attractor information dimension $\dsc$.

\begin{figure*}
\centering
\includegraphics[width=.9\textwidth]{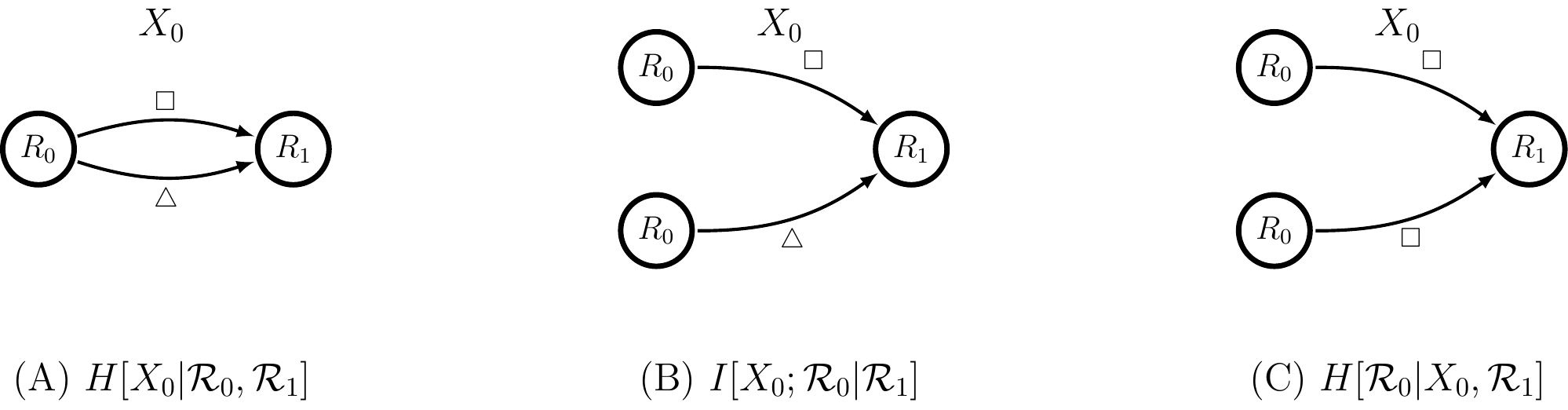}
\caption[text]{Sources of \ambiguityrate\ depicted in state machines: (A)
	$H[\MSym_0| \MxSt_0, \MxSt_{1}] > 0$---Previous state $\MxSt_0$ is mapped to
	the next state $\MxSt_1$ by two distinct symbols. This occurs when two
	symbols have identical mapping functions.
	(B) $I[\MSym_0; \MxSt_0 | \MxSt_{1}] > 0$---Two distinct previous states
	$\MxSt_0$ map to the same next state by distinct symbols, due to
	overlapping mapping functions.
	(C) $H[\MxSt_0 | \MSym_0, \MxSt_{1}] > 0$---Two distinct previous states
	$\MxSt_0$ map to the same next state by the same symbol. This occurs when a
	mapping function is noninvertible.
  }
\label{Fig:back_ent_comps}
\end{figure*}

\subsection{The Overlap Problem}
\label{TheOverlapProblem}

The \emph{overlap problem} is a long-standing concern for iterated function
systems that arises from coinciding ranges of the symbol-labeled mapping
functions $f^{(\msym)}$. \Cref{Fig:simplex_overlap} illustrates the issue and
Ref. \cite{Jurg20c} characterized it. Specifically, to quantitatively count
system orbits we must properly monitor orbit divergence and convergence.  This
then requires distinguishing between iterated function systems that meet the
open set condition (OSC) and those that do not.

\begin{Def}
\label{Def:OSC} 
An iterated function system with mapping functions $f^{(\msym)} : \simplex^N
\to \simplex^N$ satisfies the \emph{open set condition} (OSC) if there exists
an open set $U \in \simplex^N$ such that for all $\mxst, \mxstalt \in
\simplex^N$:
\begin{align*}
  f^{\mxst}(U) \cap f^{\mxstalt}(U) = \emptyset
  ~,
\end{align*}
where $\mxst \ne \mxstalt$. IFSs that meet the OSC are \emph{nonoverlapping}.
\end{Def}

When the OSC is not met, the inequality in the $\dsc$ bound \cref{eq:DSCbound}
becomes strict. This is a consequence of using $\hmu$ as our measure of state
space expansion in \cref{eq:IFSLyapunovDimension}. The Shannon entropy rate
tracks the uncertainty in the next symbol $\msym$ given our current causal state
$\mxst_t$, averaged over the Blackwell measure. From a dynamical systems point
of view, we identify this as the typical growth rate of orbits (words) in symbol
space. 

When the OSC is met, the Shannon entropy rate also measures the typical growth
rate of orbits in the $(N-1)$-simplex. Observing $\msym$, current state
$\mxst_t$ transitions to the next state $\mxst_{t+1}$ via application of the
mapping function $\mxst_{t+1} = f^{(\msym)}(\mxst_t)$. By the OSC, this is
guaranteed to be a distinct new state---thus state sequences grow at the same
rate as words do. Then, we may use $\hmu$ to measure expansion of the state
space.

However, when the OSC is not met, it is possible for two distinct states
$\mxst_t, \mxstalt_t \in \simplex$ to map to the same next state on different
symbols, by occupying the ``overlapping region'', as depicted in
\cref{Fig:simplex_overlap}. In this case, $\mxst_{t+1} = \mxstalt_{t+1}$ has no
unique pre-image. This introduces ambiguity about the past, given knowledge of
the current state. As a consequence, the Shannon entropy rate, which tracks
uncertainty in symbol space, implies a larger expansion in state space than is
actually occurring. This indicates the need to correct $\hmu$, when determining
$\dsc$.

\section{\ambiguityrate}
\label{sec:BackEntropy}

The following introduces the \emph{\ambiguityrate} to correct
$\widetilde{\dLCE}$ of \cref{eq:IFSLyapunovDimension} from overcounting orbits.
Since the problem at hand is an overestimation in uncertainty in our state
space, we must identify and quantify mechanisms of state uncertainty reduction
when the OSC is not met. Consider that when the OSC is met, every state
$\mxst_t$ has a unique pre-image $\mxst_{t-1}$ that can only be reached via a
single, specific observed symbol. When the OSC is not met, for a subset of
$\mxst \in \MxSSet$ there is uncertainty about the previous state, the previous
symbol, or both.  Quantifying this ambiguity about the past is the goal in
constructing the \ambiguityrate\ $\ha$.

Intuitively, it would seem that generating uncertainty in reverse time is
equivalent to reduction of uncertainty in forward time. The following shows
that this is the case and that the ambiguity rate is the necessary correction
to the \IFSacronym\ dimension formula \cref{eq:IFSLyapunovDimension}.

\subsection{Sources of State Uncertainty Reduction}
\label{sec:UncertaintyReduction}

For \eMs represented as \IFSacronym s, there are three distinct mechanisms that
contribute to the \ambiguityrate , as depicted in \cref{Fig:back_ent_comps}. 

The first is \emph{identical mapping functions}, depicted in
\cref{Fig:back_ent_comps} (a). When for $\msym, \msym' \in \MeasAlphabet$,
$f^{(\msym)}(\mxst) = f^{(\msym')}(\mxst)$ for all $\mxst \in \MxSSet$, we say
that $\msym$ and $\msym'$ have identical mapping functions. In this case, the
distinction between $\msym$ and $\msym'$ is not reflected in state sequences
and produces ambiguity in the symbol sequence. We quantify this as the Shannon
entropy $H[\MSym_t | \MxSt_t, \MxSt_{t+1}]$ in our current symbol, conditioned
on the previous state and the next state.

The second is overlapping mapping functions, which motivated this investigation
and already have been defined. Their impact on the state machine is shown in
\cref{Fig:back_ent_comps} (b). In this case, two distinct symbols $\msym, \msym'
\in \MeasAlphabet$ map two distinct states $\mxst, \mxstalt \in \MxSSet$ to the
same next state. Although the previous state affects the probability
distribution over the observed symbol, the next state ``forgets'' that
distinction. This is quantified by the mutual information $I[\MSym_t; \MxSt_t |
\MxSt_{t+1}]$ shared by the current symbol and the previous state, conditioned
on the next state.

Finally, there is noninvertibility in the mapping functions. If a single
function maps distinct states $\mxst, \mxstalt \in \MxSSet$ to the same
next state, the pasts that led to $\mxst$ and $\mxstalt$ can no longer be
distinguished. \Cref{Fig:back_ent_comps} (c) shows this in general. However, it
may also be observed in $f^{(\square)}$ from \cref{Fig:SNSsimplexEmbedding},
which maps every state to $\mxst_0 =  (1,0)$. The reduction via this mechanism
is measured by the Shannon entropy $H[\MxSt_t | \MSym_t, \MxSt_{t+1}]$ in the
previous state, given our next state and current symbol.

Combining these three sources of uncertainty reduction defines the
\emph{\ambiguityrate}: 
\begin{align}
  \ha = & \: H[\MSym_t| \MxSt_t, \MxSt_{t+1}] 
  + I[\MSym_t; \MxSt_t | \MxSt_{t+1}] \nonumber \\ 
  & \: \: \: \: + H[\MxSt_t | \MSym_t, \MxSt_{t+1}] \nonumber \\
  = & \: H[\MSym_t, \MxSt_t | \MxSt_{t+1}]  ~. 
\label{eq:BackEntropyInfoTheory}
\end{align}
This can be rewritten as a integral over $\MxSSet$:
\begin{align}
  \ha = - \int_{\mxst \in \MxSSet} \!\!\! \text{d} \BlackwellMeasure(\mxst) 
  \!\!\!
  \!\!\!
  \!\!\!
  \sum_{\substack{ \msym \in \MeasAlphabet , \\ \mxstalt \in (f^{(\msym)})^{-1} (\mxst)} } 
  \!\!\!
  \!\!\!
  \!\!\!
  \Pr( \msym, \mxstalt | \mxst )  \log_2 \Pr( \msym, \mxstalt | \mxst )  
  . 
\label{eq:BackEntropySum}
\end{align}
In this, we must be careful about the pre-images of $\mxst$, due to the
possibility of noninvertible mapping functions. The probability distribution
inside the summation is given by the relationship:
\begin{align}
  \Pr(\MSym_0 = \msym, \MxSt_0 = \mxstalt & | \MxSt_1 = \mxst ) = \nonumber \\
   \frac{\BlackwellMeasure( \MxSt_0 = \mxstalt )}{\BlackwellMeasure(\MxSt_1 = \mxst)} &
    \times \Pr( \MSym_0 = \msym | \MxSt_0 = \mxstalt ) ~.
\label{eq:reverseProbabilityTransition}
\end{align}
Calculating this distribution requires calculating or estimating the Blackwell
measure, which may be nontrivial.
\Cref{sec:Examples} discusses this in greater depth.

\subsection{Correcting $\dsc$}
\label{sec:Correction}

The information-theoretic decomposition of \ambiguityrate\ facilitates
combining $\ha$ and $\hmu$. Recall that for prediction, the states of a
predictive model are equivalent to knowledge of the infinite past. Due to this,
the Shannon entropy rate may be written $H[\MSym_t | \MxSt_t]$.  Combining this
with the \ambiguityrate\ gives: 
\begin{align*}
  \hmu - \ha & = H[\MSym_t | \MxSt_t] - H[\MSym_t, \MxSt_t | \MxSt_{t+1}] \\
  & = H[\MxSt_{t+1} | \MxSt_t, \MSym_t] + H[\MxSt_{t+1}] - H[\MxSt_t] \\ 
  & = \Delta H[\MxSt_t]
  ~.
\end{align*}
Moving to the third line---that is, noting that the first term
vanishes---called on the fact that the symbol and state transitions are defined
by functions. So, the difference between the Shannon entropy rate and the
\ambiguityrate\ is the growth rate of the causal state set $\MxSt$.

Recall that the information dimension, as defined in
\cref{eq:InformationDimension}, compares the average growth of occupied cells
$\mathcal{F}$---taking into account the measure over those cells---as the cell
size $\epsilon$ shrinks. To adhere to the main development, here we will not
walk through the heuristic for how a dimensional quantity is determined from
the $\LCESpectrum$. (Though, this is briefly discussed in
\cref{sec:Calculatingdsc}.) Nonetheless, we will show how the relationship
between $\di$, $\hmu-\ha$, and $\LCESpectrum$ is intuitive for \IFSacronym s in
one dimension.

When the \IFSacronym\ states lie in the 1-simplex, $\LCESpectrum$ consists of
only one exponent $\LCE_1 < 0$, which is the weighted average of the Lyapunov
exponents of each map:
\begin{align*}
\LCE_1 = \int \sum_{\msym} p^{(\msym)}(\mxst)
  \log \left| \frac{ d f^{(\msym)} (\mxst) }{d \mxst } \right| d \MxSMeasure
  ~,
\end{align*}
where $\MxSMeasure$ is the Blackwell measure.

Now, consider a line segment in $\simplex^1$ of length $\epsilon$. Mapping this
line forward $k$ times by the \IFSacronym\ produces, averaging over several
iterations of this action, $2^{(\hmu - \ha)k}$ new lines of length $\epsilon
e^{\LCE_1 k} < \epsilon$. (Note that the use of base-two for Shannon entropy
rather than base $e$ follows convention; retained here for familiarity. When
numerically estimating $\dsc$, we recommend a consistent base be chosen for
$\hmu$, $\ha$, and the $\LCESpectrum$.) The logarithmic ratio of the growth
rate of lines (as averaged over the Blackwell measure) compared to the
shrinking of these lines is the simple ratio:
\begin{align*}
  \dsc 
  & = - \frac{\hmu - \ha}{\lambda_1} ~. 
\end{align*}
This, of course, is exactly the definition of the information dimension
\cref{eq:InformationDimension} and is, assuming the \IFSacronym\ is an \eM , the
statistical complexity dimension $\dsc$.

For higher-dimensional \IFSacronym s, we conjecture that the \ambiguityrate\ is
the adjustment to the IFS Lyapunov dimension formula that gives the information
dimension: 
\begin{align}
  \widetilde{\dsc} = \left\{
    \begin{aligned}
    & k + \frac{\Lambda (k) + \hmu - \ha}{|\lambda_{k+1}|} ,
		& - \Lambda (N) > \hmu - \ha \\
    & N ~,
    & - \Lambda (N) \leq \hmu - \ha
    \end{aligned}
    \right.
    ,
  \label{eq:IFSdsc}
\end{align}
where as in \cref{eq:IFSLyapunovDimension}, $\Lambda(m)$ is the Lyapunov
spectrum partial sum $\Lambda (m) = \sum_{i=1}^m \lambda_i$ and $k = 0, 1, 2,
\ldots, N-1$ is the largest index for which $- \Lambda (k) < \hmu - \ha $.

\subsection{Interpreting Ambiguity Rate}
\label{sec:interpretingAmbiguityRate}

Up to this point, we motivated \ambiguityrate\ as correcting over counting in
the \IFSacronym\ statistical complexity dimension $\dsc$. It is worth
discussing the quantity in more depth.

On the one hand, note that when $\hmu - \ha = 0$, the causal-state process is
stationary and $\Cmu$ time-independent: $\Delta H[\MxSt_t] = 0$. This occurs
for finite-state \IFSacronym s, as well as many with countably-infinite states;
see \cref{sec:exampleSNS}. When this occurs, applying \cref{eq:IFSdsc} returns
a vanishing statistical complexity dimension $\dsc = 0$, as expected.

On the other hand, when \ambiguityrate\ vanishes, $\Cmu$ grows at the Shannon
entropy rate: $\Delta H[\MxSt_t] = \hmu$. This occurs when there are no
identical maps, no overlap, and no noninvertibility in the mapping functions.
In short, $\ha = 0$ when the causal-state process is ``perfectly self-similar''
and every new observed symbol produces a new, distinct state.

With this in mind, we can use the \ambiguityrate , and specifically $\hmu -
\ha$, to describe the stationarity of the model's internal state process. The
state set is time independent. When $\ha > 0$, however, to optimally predict
the process $\Process$ requires a nonstationary model (temporally-growing state
set $\MxSSet$), even though $\Process$ is itself stationary. This is a
consequence of modeling ``out of class''. That is, predicting a perfectly
self-similar $\Process$ requires differentiating every possible infinite past.
This is only possible with a \IFSacronym\ by storing new states at the rate new
pasts are being created. (Moving to a more powerful model class by, say, imbuing
our states with counters or stacks, may make it possible to model $\Process$
with a stationary model.)

This perspective naturally leads to another that probes the efficacy of the
causal-state mapping. Considering the space of all possible infinite pasts
$\overleftarrow{\MSym}$, the causal-state mapping
$f_{\epsilon}(\overleftarrow{\MSym})  \rightarrowtail \MxSSet$ is defined such
that:
\begin{align*}
  f_{\epsilon}(\overleftarrow{\MSym} = \overleftarrow{\msym}) = 
  f_{\epsilon}(\overleftarrow{\MSym} = \overleftarrow{\msym'}) = \mxst_i 
  ~,
\end{align*}
if $\Pr(\MSym_{0: \ell} |\overleftarrow{\MSym} = \overleftarrow{\msym}) =
\Pr(\MSym_{0: \ell} | \overleftarrow{\MSym} = \overleftarrow{\msym'})$ for all
$\ell \in \mathbf{N}^+$. When the process is perfectly self-similar, the
causal-state mapping is one-to-one and $\ha = 0$. In this case, storing the
causal states is no better for prediction than simply tracking the space of all
pasts. (Although the causal-state set $\MxSSet$ is still informative in
characterizing how we might approximate the process with a finite state machine
\cite{Marz17a}.) The number of pasts each state ``contains'' is
stationary and given by $2^{\ha} = 1$.

In general, for a stationary process $\Process$, the average number of pasts
contained by a given causal state grows at the rate $2^{\ha}$. When the process
has a stationary state set, the number of pasts each state contains must
necessarily grow at the rate new pasts are being generated, and so $2^{\hmu} =
2^{\ha}$.

\begin{figure}
  \centering
  \includegraphics[width=0.48\textwidth]{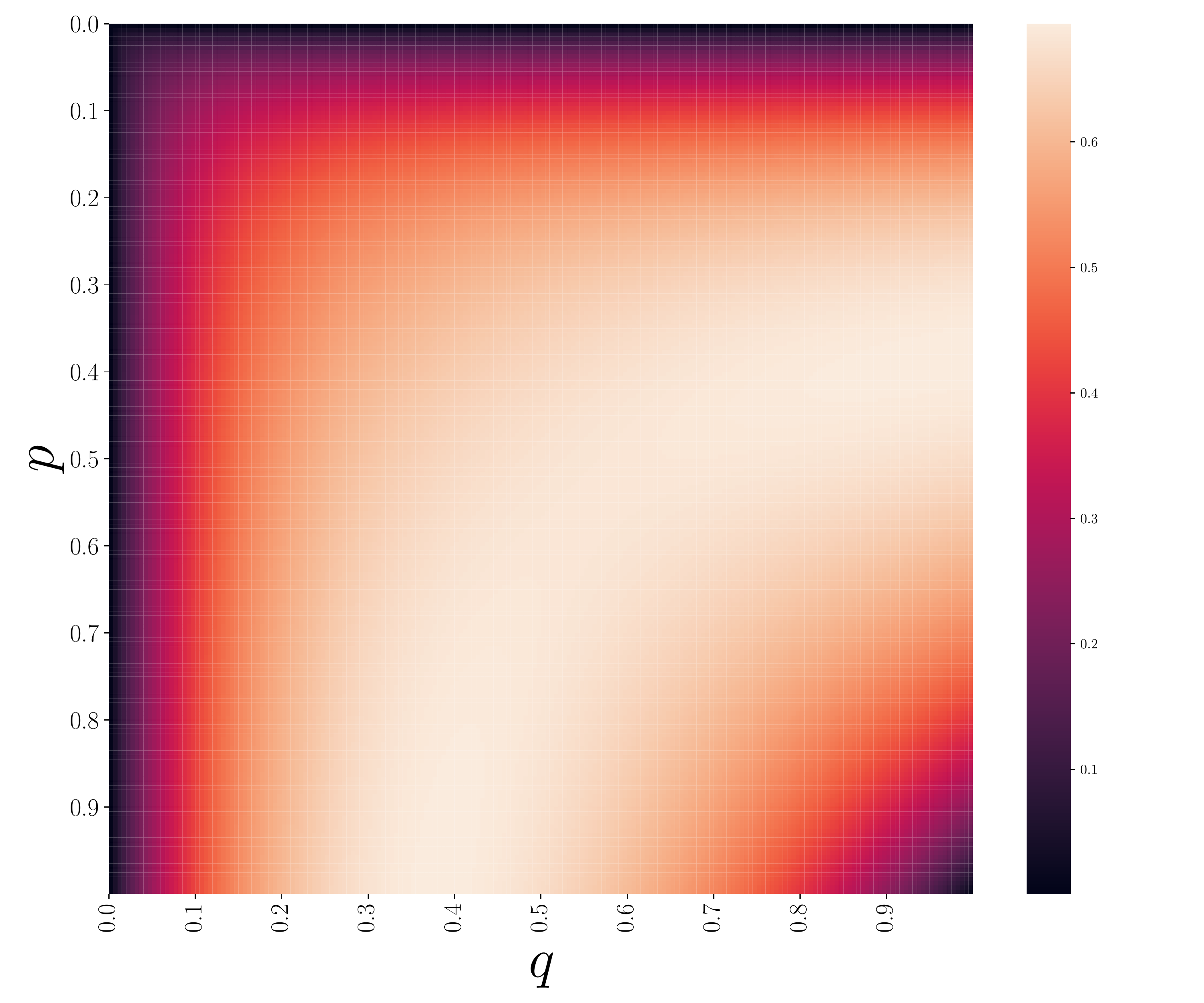}
  \caption[text]{The entropy rate $\hmu$, which in this case is equivalent to
  the \ambiguityrate\ $\ha$, is plotted for the \IFSacronym\ depicted
  in \cref{Fig:SNSsimplexEmbedding} for $p, q, \in (0,1)$. }
  \label{Fig:SNS_ent_amb}
\end{figure}

Finally, let's close with a short historical perspective. The development of
$\dsc$ was partially inspired by Shannon's definition in the 1940s of
\emph{dimension rate} \cite{Shan48a}:
\begin{align*}
\lambda = \lim_{\delta \to 0} \lim_{\epsilon \to 0} \lim_{T \to \infty}
  \frac{N ( \epsilon, \delta, T)}{T \log \epsilon}
  ~,
\end{align*}
where $N ( \epsilon, \delta, T)$ is the smallest number of elements that may be
chosen such that all elements of a trajectory ensemble generated over time $T$,
apart from a set of measure $\delta$, are within the distance $\epsilon$ of at
least one chosen trajectory. This is the minimal ``number of dimensions''
required to specify a member of a trajectory (or message) ensemble.
Unfortunately, Shannon devotes barely a paragraph to the concept, leaving it
largely unmotivated and uninterpreted. 

Therefore, it appears the first modern discussion of a dimensional quantity of
this nature for stochastic processes motivated the development using resource
theory \cite{Marz17a}, noting that the $\di$ of the causal-state set
\cref{eq:InformationDimension} characterizes the \emph{distortion rate} when
coarse-graining an uncountably-infinite state set. Starting from the
dimensional quantity, the relationship to statistical complexity was then
forged.

In this light, developing \ambiguityrate\ and calling out its easy
mathematical connection to $\Delta H[\MxSSet_t]$ flips this motivation. The
quantity $\hmu - \ha$ can be defined purely in terms of $\Process$ and has an
intuitive relationship to the causal-state mapping. The dimensional quantity
$\dsc$ naturally falls out when we compare this rate of model-state growth to
the dynamics of the causal states in the mixed-state simplex. Therefore, we may
motivate $\dsc$ not as only a resource-theoretic tool for finitizing
infinitely-complex state machines, but also as an intrinsic measurement of
a process' structural complexity.

\section{Examples}
\label{sec:Examples}

We now consider two examples. The first is a parametrized discrete-time
renewal process that has a countably-infinite state space for all parameters.
This allows us to explicitly write down the Blackwell measure and calculate
\ambiguityrate\ exactly using \cref{eq:BackEntropySum}. The second is a
parametrized machine with three maps, which has an uncountably-infinite state
space for nearly all parameters. Calculating the \ambiguityrate\ in this case
requires us to approximate the Blackwell measure using Ulam's method. 

\begin{figure*}[ht]
\centering
\includegraphics[width=\textwidth]{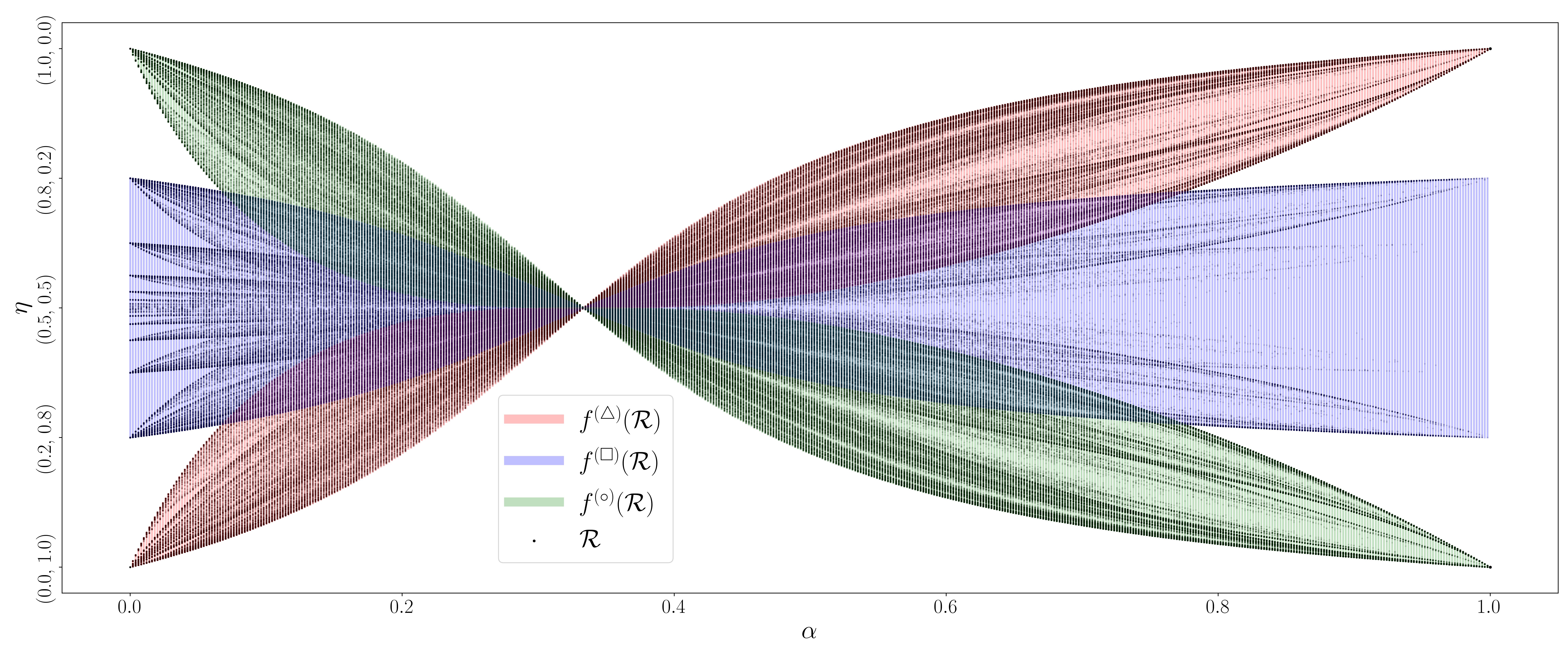}
\caption[text]{One-dimensional attractor for \IFSacronym\ given in
	\cref{eq:1D_sarah_machine} with $x =  0.25$ and horizontally varying
	$\alpha \in (0,1)$. State set $\mxst \in \MxSSet$ plotted (red, blue,
	green) on top of the images of the mapping functions
	$\left\{f^{(\triangle)}, f^{(\square)}, f^{(\circ)} \right\}$ applied to
	$\MxSSet$. For $\alpha \in (0.07, 0.78)$, there is overlap in the images of
	the maps.
	}
\label{fig:1D_param_uncount_machine}
\end{figure*}

\subsection{Example: Discrete-Time Renewal Process}
\label{sec:exampleSNS}

The \IFSacronym\ depicted in \cref{Fig:SNSsimplexEmbedding} has the alphabet
$\MeasAlphabet = \{ \triangle, \square \}$ and the substochastic matrices:
\begin{align}
  T^{\triangle} = \begin{pmatrix}
    1-q & q  \\
    0 & 1-p
\end{pmatrix}
~\text{and}~
T^{\square} = \begin{pmatrix}
  0 & 0 \\
  p & 0
\end{pmatrix}
~,
\label{eq:SNS_matrices}
\end{align}
where $p = q = \frac{1}{2}$. For the general case where $p, q \in (0, 1)$ are
left unspecified, we have the probability function set:
\begin{align*}
  p^{(\triangle)}(\mxst) &  = 1 - \langle \mxst | \delta_2 \rangle p ~, \\
  p^{(\square)}(\mxst) & = \langle \mxst | \delta_2 \rangle p ~,
\end{align*}
and the mapping function set:
\begin{align*}
  f^{(\triangle)}(\mxst) &  = \left( 
    \frac{\langle \mxst | \delta_1 \rangle (1-q) }{1 - \langle \mxst | \delta_2 \rangle p},
    \frac{\langle \mxst | \delta_1 \rangle q + \langle \mxst | \delta_2 \rangle (1-p)  }
    {1 - \langle \mxst | \delta_2 \rangle p}
    \right)~, \\
  f^{(\square)}(\mxst) & = \left( 1, 0 \right)
  ~. 
\end{align*}

Since $\MxSSet$ is countable, we may write down the state set $\MxSSet$ as a
sequence:
\begin{align*}
   \mxst_n = \left[ \frac{(p-q)(1-q)^n}{p(1-q)^n - q(1-p)^n}, 
   \frac{q(1-q)^n - q(1-p)^n}{p(1-q)^n - q(1-p)^n} \right] ~, 
\end{align*}
where $n$ is the number of $\triangle$s seen since the last $\square$ and $p
\neq q$. 
This simple structure allows us to give the Blackwell measure explicitly:
\begin{align*}
\BlackwellMeasure(n) = \frac{p(1-q)^n - q(1-p)^n}{p-q} \times \frac{pq}{p+q}
  ~,
\end{align*}
where $\BlackwellMeasure(n)$ is the asymptotic invariant measure over the state
induced after seeing $n$ $\triangle$s since the last $\square$.

With the Blackwell measure in hand, the entropy rate can be explicitly
calculated as the infinite sum:
\begin{align*}
  \hmu = & \: \sum_{n=1}^{\infty} \mu_n H [ \MSym_n | \MxSt_n = \mxst_n ] \\ 
  = & \: - \sum_{n=1}^{\infty} \mu_n \Big( p^{(\triangle)} (\mxst_n) 
  \log_2 p^{(\triangle)} (\mxst_n) \\
  & \qquad\qquad\quad  + p^{(\square)} (\mxst_n) \log_2 
  p^{(\square)} (\mxst_n) \Big)
  ~.
\end{align*}
\Cref{Fig:SNS_ent_amb} plots $\hmu$ for $p, q \in (0,1)$.
In calculating $\hmu$, there is a contribution from every state except the
first---$\mxst_0$---since the first state transitions to the second with
probability one and there is no branching uncertainty. Every other state
transitions on a coin flip of a determined bias between $(\triangle, \square)$,
generating uncertainty with each transition.

In contrast to how $\hmu$ averages over all mixed states, \ambiguityrate\
accumulates in only one state---$\mxst_0$. From \cref{Fig:SNSsimplexEmbedding},
we see that $H[\msym_n, \mxst_n | \mxst_{n+1}] = 0$ for all $n$ other than
$n=0$. That is, each state $\mxst_n$ is only accessed via the prior state
$\mxst_{n-1}$, except for $\mxst_0$, which may be accessed from every other state.
So, ambiguity in the past can only be introduced by visiting $\mxst_0$. Since
these transitions only occur on a $\square$, we must find the probability
distribution $\Pr(\MSym_0 = \square, \MxSt_0 = \mxst_n | \MxSt_1 = \mxst_0)$. 

Applying \cref{eq:BackEntropySum} and \cref{eq:reverseProbabilityTransition}, we
explicitly write down the \ambiguityrate\ as: 
\begin{align*}
  \ha = \mu_0 \sum_{n = 1}^{\infty} 
  \left( \frac{\mu_n}{\mu_0}  p^{(\square)} (\mxst_n) \right)
  \log_2 \left( \frac{\mu_n}{\mu_0}  p^{(\square)} (\mxst_n) \right)
  ~.
\end{align*}
Both $\hmu$ and $\ha$ are infinite summations, but when calculating the
\ambiguityrate , the sum refers to calculating a single Shannon entropy over the
infinite, discrete distribution representing the probability distribution over
prior states when arriving in $\mxst_0$. 

Since the state space does not grow---$\Delta H[\MxSSet_t] = 0$---the entropy
rate $\hmu = \ha$ as $n\to\infty$. Therefore, $\dsc$ vanishes for all values of
$p$ and $q$. This will always be the case for finite-state \IFSacronym s and, in
general, for those with countable state spaces.

\begin{figure*}
\centering
\includegraphics[width=\textwidth]{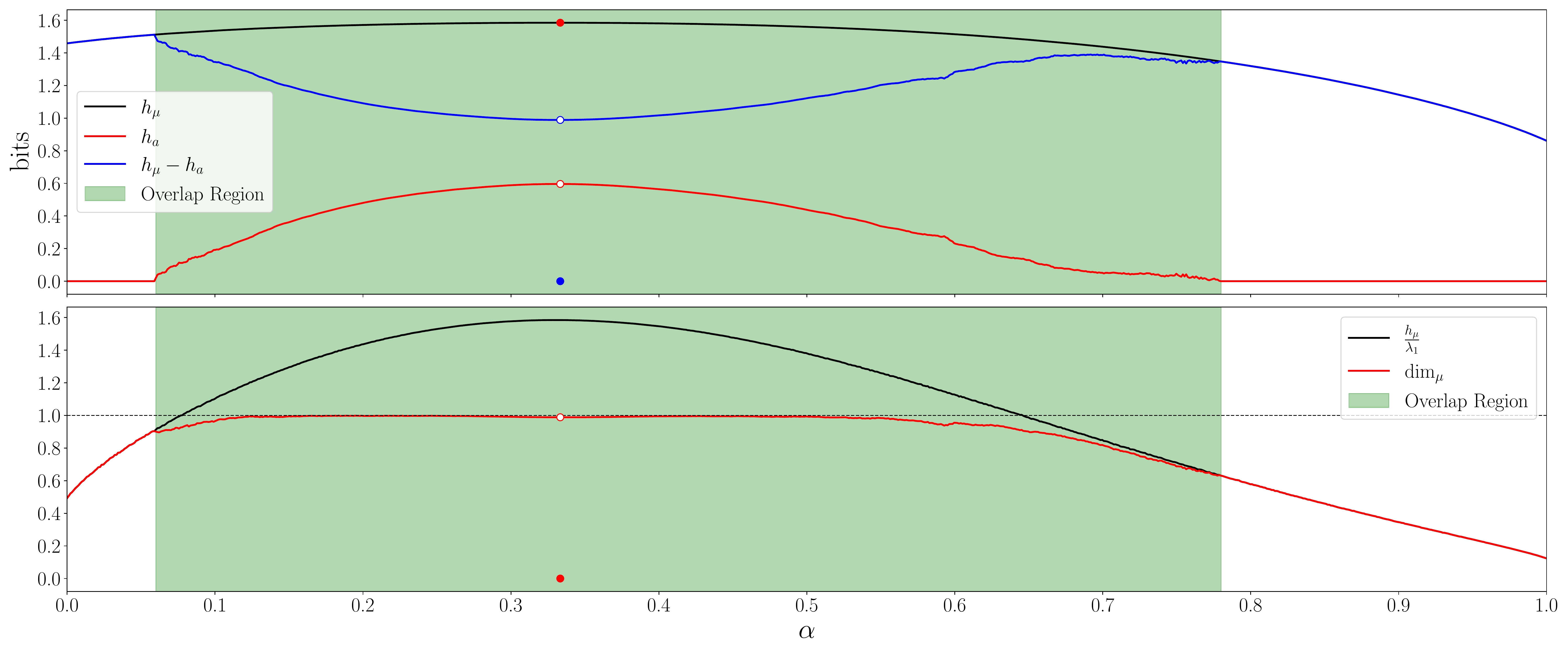}
\caption[text]{Calculating entropies and dimensions for the \IFSacronym\ given
	in \cref{eq:1D_sarah_machine} with $\alpha \in (0,1)$, and $x =  0.25$:
	(Top) The entropy rate $\hmu$, the ambiguity rate $\ha$, and $\hmu-\ha$.
	(Bottom) Comparing $\hmu / \lambda_1$ to $\dsc = (\hmu - \ha) / \lambda_1$:
	The latter smoothly departs from the former and is approximately $1$ for
	much of the overlap region, except where it discontinuously jumps to zero
	at $\alpha = 1/3$.
	}
\label{fig:1D_param_uncount_machine_ents}
\end{figure*}

\subsection{Example: 1-D Ambiguity Rate}
\label{sec:OneDimAmbiguityRate}

Now, let's turn to the more general case, those \IFSacronym s with
uncountably-infinite state spaces. For the moment we restrict to
one-dimensional \IFSacronym s, so that the states lie in the $1$-simplex.
Consider a \IFSacronym\ with the alphabet $\MeasAlphabet = \left\{ \triangle,
\square, \circ \right\}$ and the associated substochastic matrices: 
\begin{align}
  \label{eq:1D_sarah_machine}
  T^{\triangle} & = \begin{pmatrix}
          \alpha y & \beta x \\
          \alpha x & \beta y 
      \end{pmatrix}
      ,~
  T^{\square} = \begin{pmatrix}
          \beta y  & \beta x \\
          \beta x  & \beta y    
      \end{pmatrix}
      ,~\text{and}
    \nonumber
      \\
  T^{\circ} & = \begin{pmatrix}
          \beta y & \alpha x \\
          \beta x & \alpha y 
      \end{pmatrix}
      ~,
\end{align}
with $\alpha = 1 - 2 \beta, \alpha \in (0,1)$, and $x = 1-y, x\in (0,1)$. 

\Cref{fig:1D_param_uncount_machine} depicts all of the \IFSacronym s for the
slice of the parameter space where $x=0.25$. The vertical axis is the
$1$-simplex and each vertical slice plots the state space $\MxSSet(\alpha)$ at
the appropriate value of $\alpha$, given on the horizontal axis. Additionally,
the images of the functions $\left\{f^{(\triangle)}(\MxSSet),
f^{(\square)}(\MxSSet), f^{(\circ)}(\MxSSet) \right\}$ are shaded in red, blue,
and green, respectively.

At $\alpha = 1/3$, the mixed state set $\MxSSet$ contracts to a finite set, and
$\hmu$ must equal to $\ha$, making $\dsc = 0$. At this point in parameter space,
the state set consists of only one state; $\MxSSet(\alpha = 1/3) = \{(1/2,
1/2)\}$. At every other value of $\alpha$, $\ha < \hmu$. There is overlap in the
images of the maps for, approximately, $\alpha \in (0.07, 0.78)$. In this
regime, $\ha > 0$.

To calculate the ambiguity rate and therefore the statistical complexity
dimension $\dsc$ we use a modified Ulam's method to approximate the Blackwell
measure and then approximate the integral equation \cref{eq:BackEntropySum}.
This method is not the only way to find the ambiguity rate, but does have
several advantages, including speed and the ability to control the accuracy of
our approximation. This method is discussed in depth in
\cref{app:NumericalCalculation}.

The top plot in \cref{fig:1D_param_uncount_machine_ents} gives the entropy rate,
ambiguity rate, and $\hmu - \ha$ for the \IFSacronym s pictured in
\cref{fig:1D_param_uncount_machine}. As $\alpha$ is increased, $\ha$ smoothly
increases from zero as overlap begins to occur. It approaches $\approx 0.6$
around $\alpha = 1/3$, but is discontinuously equal to zero at this point. The
reason for this is an instantaneous equality in the fixed points of the mapping
functions, causing the state space to collapse. As $\alpha$ increases to $1$,
$\ha$ smoothly decreases back to zero. The roughness seen in the plot is due to
numerical precision, as explained in \cref{app:NumericalCalculation}. 

For a large portion of the overlap region, $\dLCE$ saturates at 1.0. The bottom
plot in \cref{fig:1D_param_uncount_machine_ents} instead depicts $\hmu /
\lambda_1$ to show how this quantity smoothly changes across parameter space,
reaching a maximum around $\alpha \approx 1.6$. By way of comparison, $\dsc =
(\hmu - \ha) / \lambda_1$ smoothly departs from the Lyapunov dimension when
overlap begins and, instead, asymptotes from below to the $\dim = 1.0$ line for
much of the overlap region. Again, at $\alpha = 1/3$ there is the discontinuous
drop to $\dsc=0$, followed by $\dsc$ smoothly rejoining with the Lyapunov
dimension as the overlap region ends. 

Unsurprisingly, calculating the \ambiguityrate\ in higher dimensions is more
challenging. Although, in principle, Ulam's method still applies and we may in
principle follow the algorithm laid out in \cref{app:NumericalCalculation},
higher-dimensional mapping functions introduce additional error sources in the
approximation. Developing an algorithm to efficiently and accurately calculate
the \ambiguityrate\ in higher dimensions is of great interest. We leave this
task to future work, however, having achieved our goal of introducing a method
to estimate the statistical complexity dimension. 

\section{Conclusion}
\label{sec:Conclusion}

Stepping back from developing ambiguity rate and statistical complexity
dimension, let us position the new results here in the context of our prior two
works in this series \cite{Jurg20b, Jurg20c}. In the first, motivated by
needing a general solution to the Shannon entropy rate for processes generated
by finite-state hidden Markov chains, we showed how an optimal predictor can be
constructed for any such process, at the cost of a potentially
uncountably-infinite state space. To address the resulting challenge, we
introduced hidden Markov-driven iterated function systems and showed that the
attractor of a properly-defined \IFSacronym\ is equivalent to the \eM for the
process generated by its substochastic matrices. 

The result gave benefits beyond a finite-dimensional description of an
infinite-state model. The identification allowed us to adopt several rigorous
results on IFSs, including an ergodic theorem that allows us to sample the
\IFSacronym\ to accurately and efficiently calculate the Shannon entropy rate
of the underlying process. With this, our original goal was completed.

However, identifying these \eMs as IFSs allowed us to show that the dimension
of the mixed state set, a quantity well studied for IFSs, is a structural
complexity measure for stochastic processes. The second prequel Ref.
\cite{Jurg20c} then introduced the statistical complexity dimension---the
\IFSacronym\ attractor information dimension. A long-standing conjecture in
dynamical systems theory states that the Lyapunov dimension, a dimensional
quantity calculated using a system's Lyapunov spectrum, is equivalent to the
information dimension. We showed that for many \IFSacronym s this is indeed the
case, connecting the information dimension of the \eM's state space to the
\eM's statistical complexity dimension---the rate of divergence of the
statistical complexity. This related a \IFSacronym's dynamics to the
information-theoretic properties of the underlying process. Additionally, it
gave a new and meaningful measure of structural complexity---one that
differentiates between stochastic processes with divergent state spaces.

That was not the end of the story, since calculating $\dsc$ is difficult due to
long-standing challenges in the field of IFS dimension theory. In particular,
the overlap problem posed a significant hurdle---restricting the preceding
results to only nonoverlapping IFSs. This limited analyses to the class of
stochastic processes with one-to-one past-to-causal state mappings. In one
sense, these processes are the most complex but exhibit structure that is the
least interesting. That is, for processes generated by these \IFSacronyms\ one
simply stores every past to build an optimally-predictive model.

This state of affairs led directly to the present development and to
introducing the ambiguity rate. The latter allows smoothly varying between \eMs
with countable state spaces ($\ha = \hmu$ and $\Delta H[\MxSSet]  = 0$) and
those with perfectly self-similar state spaces ($\ha = 0$ and $\Delta
H[\MxSSet] = \hmu$), including all those lying in between, with $ \hmu > \ha >
0$ and $\Delta H[\MxSSet] = \hmu - \ha$. This model class is much more general,
generating an exponentially larger family of stochastic processes. As such, we
anticipate that this class will be of great interest and likely to lead to
significant further progress in analyzing the randomness and structure
generated by hidden Markov chains.

To close, we note that the structural tools and the entropy-rate method
introduced by this trilogy were put to practical application in two other
previous works. One diagnosed the origin of randomness and structural
complexity in quantum measurement \cite{Vene19a}. The other exactly determined
the thermodynamic functioning of Maxwellian information engines \cite{Jurg20a},
when there had been no previous method for this kind of detailed and accurate
classification. The lesson from these applications of finite-state-generated
processes is that the resulting effectively-infinite state processes are very
likely generic. That said, for now we must leave to the future investigating
infinite-state machines and developing the required algorithmic tools.

\section*{Acknowledgments}
\label{sec:acknowledgments}

The authors thank Alec Boyd, Sam Loomis, and Ryan James for helpful discussions
and the Telluride Science Research Center for hospitality during visits and the
participants of the Information Engines Workshops there. JPC acknowledges the
kind hospitality of the Santa Fe Institute, Institute for Advanced Study at the
University of Amsterdam, and California Institute of Technology for their
hospitality during visits. This material is based upon work supported by, or in
part by, FQXi Grant number FQXi-RFP-IPW-1902, and U.S. Army Research Laboratory
and the U.S. Army Research Office under grants W911NF-18-1-0028 and W911NF-
21-1-0048.

\appendix

\onecolumngrid
\clearpage
\begin{center}
{\huge Supplementary Materials}\\
\vspace{0.1in}
\vspace{0.1in}
{\huge Ambiguity Rate of\\[10pt]
Hidden Markov Processes}\\[15pt]
{\large Alexandra Jurgens and James P. Crutchfield\\[5pt]
\arxiv{2002.XXXXX}
}
\end{center}

\setcounter{equation}{0}
\setcounter{figure}{0}
\setcounter{table}{0}
\setcounter{page}{1}
\setcounter{section}{0}
\makeatletter
\renewcommand{\theequation}{S\arabic{equation}}
\renewcommand{\thefigure}{S\arabic{figure}}
\renewcommand{\thetable}{S\arabic{table}}

The Supplementary Materials to follow give a suite of example hidden Markov
chains and discuss numerically estimating the ambiguity rate.

\section{Hidden Markov-Driven Iterated Function System Examples}
\label{app:nonunifilarHMCs}

We reproduce here the hidden Markov-driven iterated function systems
(\IFSacronym) used to create \cref{fig:simplex_examples}. 

First, the \emph{delta} \IFSacronym, from \cref{fig:simplex_example_delta}, is
given by a three-symbol alphabet and the substochastic symbol-labeled matrices:
\begin{align}
  \label{eq:delta_machine}
  T^{\square} & = \begin{pmatrix}
    0.112 & 0.355 & 3.901 \times 10^{-2} \\
    0.434 & 7.685 \times 10^{-2} & 2.333 \times 10^{-2} \\
    0.215 & 2.518 \times 10^{-2} & 0.220
      \end{pmatrix}
      , \\
  T^{\triangle} & = \begin{pmatrix}
    1.778 \times 10^{-2} & 0.113 & 0.220 \\
    6.465 \times 10^{-2} & 0.272 & 2.413 \times 10^{-2} \\
    0.400 & 8.697 \times 10^{-3}  & 9.892 \times 10^{-3}        
      \end{pmatrix}
      ,~\text{and}~
    \nonumber
      \\
  T^{\circ} & = \begin{pmatrix}
    8.312 \times 10^{-2} & 2.867 \times 10^{-2} & 3.096 \times 10^{-2} \\
    4.690 \times 10^{-2} & 5.625 \times 10^{-2}  &  1.807 \times 10^{-3} \\
    0.114 & 1.095 \times 10^{-3} & 7.522 \times 10^{-4}       
      \end{pmatrix}
      ~,
    \nonumber
\end{align}

Second, the \emph{Nemo} \IFSacronym, from \cref{fig:simplex_example_nemo}, is
given by a two-symbol alphabet and the substochastic symbol-labeled matrices:
\begin{align}
  \label{eq:nemo_machine}
  T^{\square} & = \begin{pmatrix}
    0.409 & 0.0   & 0.091 \\
    0.5   & 0.0 & 0.0 \\
    0.0   & 0.182 & 0.0
      \end{pmatrix}
      ,~\text{and}~
       \\
  T^{\triangle} & = \begin{pmatrix}
    0.091 & 0.0   & 0.409 \\
    0.5   & 0.0 & 0.0 \\
    0.0   & 0.818 & 0.0    
      \end{pmatrix}
      ~,
    \nonumber
\end{align}

Finally, the \emph{gamma} \IFSacronym , from \cref{fig:simplex_example_gamma},
is given by a three-symbol alphabet and the substochastic symbol-labeled
matrices:
\begin{align}
  \label{eq:gamma_machine}
  T^{\square} & = \begin{pmatrix}
    2.479 \times 10^{-2} & 0.355 & 1.745 \times 10^{-2} \\
    0.410 & 1.878 \times 10^{-2} & 2.388 \times 10^{-4} \\
    0.204 & 2.472 \times 10^{-3} & 0.215
      \end{pmatrix}
      , \\
  T^{\triangle} & = \begin{pmatrix}
    1.672 \times 10^{-3} & 0.133 & 0.235 \\
    3.377 \times 10^{-2} & 0.272 & 8.277 \times 10^{-2} \\
    0.426 & 1.498 \times 10^{-2}  & 4.286 \times 10^{-3}        
      \end{pmatrix}
      ,~\text{and}~
    \nonumber
      \\
  T^{\circ} & = \begin{pmatrix}
    8.870 \times 10^{-2} & 3.059 \times 10^{-2} & 0.114 \\
    6.918 \times 10^{-2} & 0.112  &  1.804 \times 10^{-3} \\
    0.131 & 1.165 \times 10^{-3} & 8.005 \times 10^{-4}       
      \end{pmatrix}
      ~,
    \nonumber
\end{align}

Due to finite numerical accuracy, reproducing the attractors using these
specifications may differ slightly from \cref{fig:simplex_examples}. 

The mapping images shown in \cref{Fig:simplex_overlap} are produced by the
following three-symbol \IFSacronym:
\begin{align}
  \label{eq:sarah_machine}
  T^{\square} & = \begin{pmatrix}
          \alpha y & \beta x & \beta x \\
          \alpha x & \beta y & \beta x \\
          \alpha x & \beta x & \beta y
      \end{pmatrix}
      ,~
  T^{\triangle} = \begin{pmatrix}
          \beta y & \alpha x & \beta x \\
          \beta x & \alpha y & \beta x \\
          \beta x & \alpha x & \beta y        
      \end{pmatrix}
      ,~\text{and}
    \nonumber
      \\
  T^{\circ} & = \begin{pmatrix}
          \beta y & \beta x & \alpha x \\
          \beta x & \beta y & \alpha x \\
          \beta x & \beta x & \alpha y        
      \end{pmatrix}
      ~,
\end{align}
with $\alpha = 0.63$ and $x=0.2$ for the overlapping example in
\cref{fig:simplex_example_nonoverlapping} and $\alpha = 0.6$ and $x= 0.15$ for
the nonoverlapping example in \cref{fig:simplex_example_overlapping}.

\begin{figure*}
  \centering
  \includegraphics[width=\textwidth]{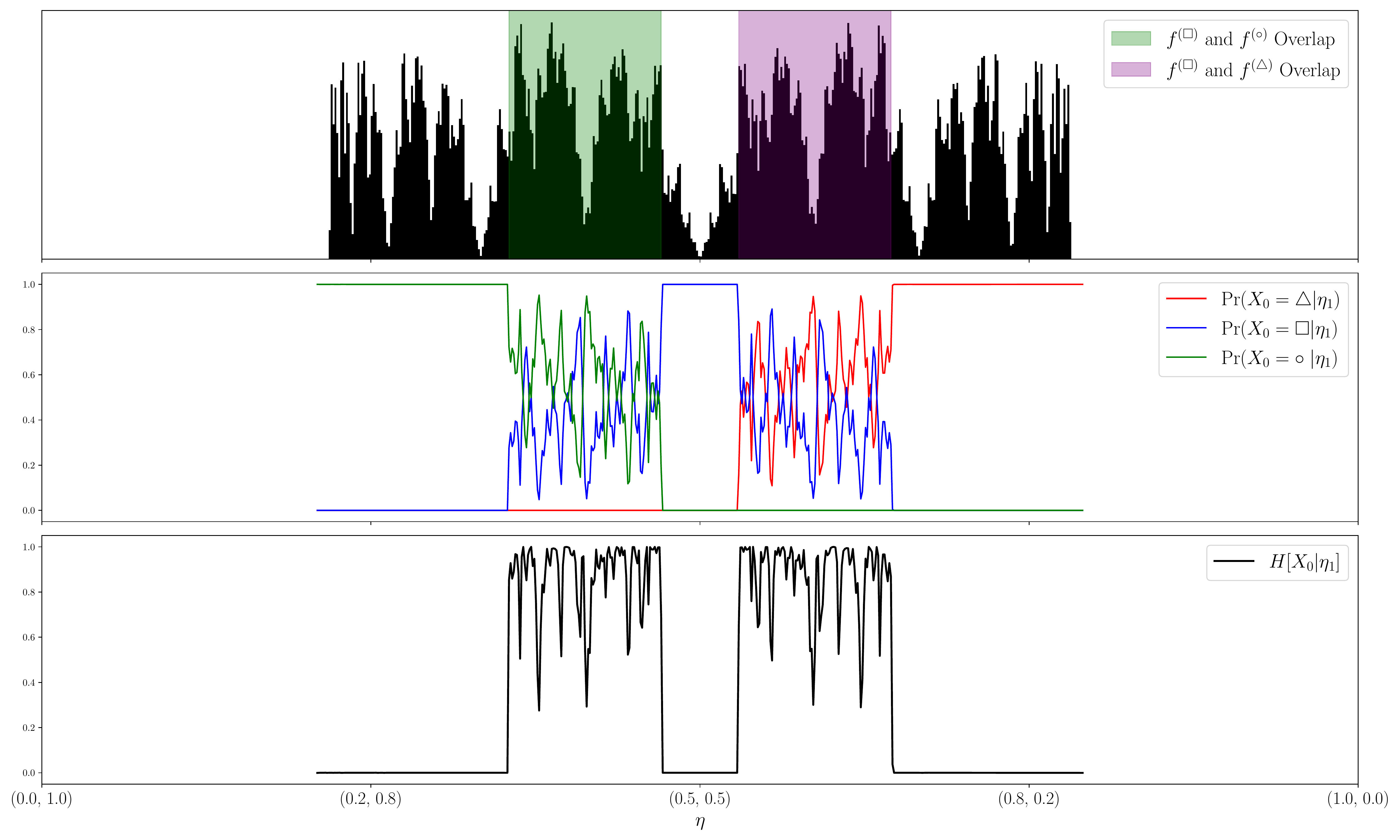}
  \caption[text]{\IFSacronym\ for $x = 0.25$ and $\alpha=0.5$: (Top) Blackwell
    Measure $\BlackwellMeasure$ approximated by Ulam's method with $k = 400$.
    The two overlapping regions are overlaid and may be compared with
    \cref{fig:1D_param_uncount_machine}. (Middle) Probability of each prior map
    is plotted. In nonoverlapping regions, only one prior map is possible. In
    the overlapping regions, there are complicated, fractal-like distributions
    over multiple prior maps. (Bottom) Shannon entropy over the prior map:
    Nonzero only in the overlapping regions.
    }
  \label{fig:1D_param_uncount_machine_plots}
\end{figure*}

\section{Numerical Approximation of Ambiguity Rate}
\label{app:NumericalCalculation}

To estimate the ambiguity rate for a \IFSacronym\ lying in the $1$-simplex, we
may use Ulam's method to approximate the Blackwell measure, then compute
\cref{eq:BackEntropySum}. Given a partition $\{ A_1, \dots A_k\}$ of the
simplex, define:
\begin{align*}
P_{ij}^{(\msym)} = \frac{m(f^{(\msym)} (A_i) \cap A_j)}{m(f^{(\msym)}(A_i))}
  \times p^{(\msym)} \left( \overline{A_i} \right)
  ~,
\end{align*}
where $m$ is the Lebesgue measure over $\simplex$ and $\overline{A_i}$ is the
center of a partition element. Let $P = \sum P^{(\msym)} $ and find the left
eigenvalue $p = pP$. Then, the invariant-measure approximation is:
\begin{align*}
  \mu_n(A) = \sum_i p_i \frac{m(A\cap A_i)}{m(A_i)}
  ~.
\end{align*}

For this example, let's walk through estimating the \ambiguityrate\ for one
\IFSacronym---setting $x=0.25$ and $\alpha =0.5$. The partition $\{ A_1, \dots
A_k\}$ is created by dividing the $1$-simplex into $k$ boxes of equal length.
The approximated Blackwell measure $\widehat{\BlackwellMeasure}$ for the
\IFSacronym , using $k = 400$, is shown in the top plot of
\cref{fig:1D_param_uncount_machine_plots}. The overlay indicates the region
(green) of the state space that exhibits overlap. Compare the two regions
depicted in \cref{fig:1D_param_uncount_machine_plots} to the overlap shown in
\cref{fig:1D_param_uncount_machine}, for the vertical slice at $\alpha = 0.5$.

Note that the partition may be defined as desired. We have found that defining
the partition by calculating the set of fixed points of the mapping functions
$\left\{ p^{\msym} : f^{(\msym)} (p^{\msym}) = p^{\msym} \right\}$. Then, as
many times as is desired, find all possible iterates of each fixed point,
constructing a new set $\left\{ f^{(w)} (p^{\msym}) : \msym  \in \MeasAlphabet,
w \in \bigcup_{n = 0}^{N} \MeasAlphabet^n \right \}$, where $N \in
\mathbb{Z}^+$. Removing duplicates and ordering the set gives a list of
endpoints for a partition of the $1$-simplex. Increasing $N$ produces
increasingly fine partitions. This method of defining partitions has advantages
when calculating $\ha$ across parameter space as we have in
\cref{sec:OneDimAmbiguityRate}, since the position of the fixed point iterates
in the simplex are smooth functions of $\alpha$. 

Regardless, once the partition is selected and $\widehat{\BlackwellMeasure}$ is
determined, we again use the partition. For each cell $A_i$, we find the
probability distribution over the maps that could have transitioned into $A_i$.
By applying \cref{eq:reverseProbabilityTransition} and assuming invertibility of
the mapping functions gives:
\begin{align*}
\Pr(\MSym_0 = \msym | \MxSt \in A_i )
  = & \: \frac{\widehat{\BlackwellMeasure} 
  \left( \left( f^{(\msym)} \right)^{-1} (A_i)\right)  }{ \widehat{\BlackwellMeasure} (A_i )} 
  ~p^{\msym} \left( \overline{\left( f^{(\msym)} \right)^{-1} (A_i)} \right) 
  ~.
\end{align*}

For our example \IFSacronym , the probability of the previous map given current
location in the simplex is plotted in the middle figure of \cref{fig:1D_param_uncount_machine_plots}. For parts of the
simplex outside the overlapping regions, only one prior map is possible and it has probability one. Within the overlapping regions, the distribution over the
possible prior maps may be very complicated. The Shannon entropy over the prior
map distribution $H \left[ \MSym_0 = \msym | \MxSt \in A_i  \right]$ is shown in
the third plot of \cref{fig:1D_param_uncount_machine_plots}. Once these
entropies are calculated, the final step is to approximate the integral equation
\cref{eq:BackEntropySum} with a summation over cells in the partition:
\begin{align*}
  \ha = \sum_i \widehat{\BlackwellMeasure} ( A_i ) 
  \sum_{\msym \in \MeasAlphabet} H \left[ \MSym_0 = \msym | \MxSt \in A_i  \right] ~.
\end{align*} 
In our example, the \ambiguityrate\ is found to be $\ha = 0.4499$. Since the
\IFSacronym\ entropy rate is $\hmu = 1.5596$, this gives an adjusted state
space expansion rate of $\hmu - \ha = 1.1098$. Calculating the \IFSacronym 's
$\LCESpectrum$ and applying \cref{eq:IFSdsc} results in a statistical
complexity dimension of $\dsc = 0.9815$.

The advantage of Ulam's method is its relative simplicity and computational
speed. Additionally, it is deterministic given the partition. And, we may may
increase estimation accuracy simply by tuning our partition; although,
increasingly-fine partitions increase computation time.

Additionally, when the set becomes highly rarefied, fluctuations will be
observed in the $\ha$ estimates. This can be seen in our example \IFSacronym\
at either end of the overlap region; although, it is worst when $\alpha \in
(0.6, 0.78)$. This may be understood when comparing
\cref{fig:1D_param_uncount_machine} to
\cref{fig:1D_param_uncount_machine_ents}. From $\alpha \in (0.6, 0.78)$ there
are bands of high density in the overlapping region that increase in
probability as the overlapping region itself shrinks. Calculating $\ha$
accurately in this region requires increasingly-fine partitioning. An immediate
improvement may be made by changing the method to use adaptive partitioning
while sweeping parameter space. This adapts to the changing structure of the
state set. The method may be applied to any \IFSacronym\ in the $1$-simplex
with overlaps. 


\begin{thebibliography}{10}

\bibitem{Turi37a}
A.~Turing.
\newblock On computable numbers, with an application to the
  {Entschiedungsproblem}.
\newblock {\em Proc. Lond. Math. Soc.}, 42, 43:230--265, 544--546, 1937.

\bibitem{Shan56c}
C.~E. Shannon.
\newblock A universal {Turing} machine with two internal states.
\newblock In C.~E. Shannon and J.~McCarthy, editors, {\em Automata Studies},
  number~34 in Annals of Mathematical Studies, pages 157--165. Princeton
  University Press, Princeton, New Jersey, 1956.

\bibitem{Mins67}
M.~Minsky.
\newblock {\em Computation: Finite and Infinite Machines}.
\newblock Prentice-Hall, Englewood Cliffs, New Jersey, 1967.

\bibitem{Shan48a}
C.~E. Shannon.
\newblock A mathematical theory of communication.
\newblock {\em Bell Sys. Tech. J.}, 27:379--423, 623--656, 1948.

\bibitem{Kolm56b}
A.~N. Kolmogorov.
\newblock {\em Foundations of the Theory of Probability}.
\newblock Chelsea Publishing Company, New York, second edition, 1956.

\bibitem{Kolm65}
A.~N. Kolmogorov.
\newblock Three approaches to the concept of the amount of information.
\newblock {\em Prob. Info. Trans.}, 1:1, 1965.

\bibitem{Kolm83}
A.~N. Kolmogorov.
\newblock Combinatorial foundations of information theory and the calculus of
  probabilities.
\newblock {\em Russ. Math. Surveys}, 38:29--40, 1983.

\bibitem{Kolm59}
A.~N. Kolmogorov.
\newblock Entropy per unit time as a metric invariant of automorphisms.
\newblock {\em Dokl. Akad. Nauk. SSSR}, 124:754, 1959.
\newblock (Russian) Math. Rev. vol. 21, no. 2035b.

\bibitem{Sina59}
Ja.~G. Sinai.
\newblock On the notion of entropy of a dynamical system.
\newblock {\em Dokl. Akad. Nauk. SSSR}, 124:768, 1959.

\bibitem{Crut12a}
J.~P. Crutchfield.
\newblock Between order and chaos.
\newblock {\em Nature Physics}, 8(January):17--24, 2012.

\bibitem{Jurg20b}
A.~Jurgens and J.~P. Crutchfield.
\newblock Shannon entropy rate of hidden {Markov} processes.
\newblock {\em J. Statistical Physics}, to appear, 2020.
\newblock arXiv.org:2008.12886.

\bibitem{Jurg20c}
A.~Jurgens and J.~P. Crutchfield.
\newblock Divergent predictive states: The statistical complexity dimension of
  stationary, ergodic hidden {Markov} processes.
\newblock {\em arxiv.org:2102.10487}, 2021.

\bibitem{Marc11a}
B.~Marcus, K.~Petersen, and T.~Weissman, editors.
\newblock {\em Entropy of Hidden Markov Process and Connections to Dynamical
  Systems}, volume 385 of {\em Lecture Notes Series}. London Mathematical
  Society, 2011.

\bibitem{Ephr02a}
Y.~Ephraim and N.~Merhav.
\newblock Hidden {Markov} processes.
\newblock {\em IEEE Trans. Info. Th.}, 48(6):1518--1569, 2002.

\bibitem{Bech15a}
J.~Bechhoefer.
\newblock Hidden {Markov} models for stochastic thermodynamics.
\newblock {\em New. J. Phys.}, 17:075003, 2015.

\bibitem{Rabi86a}
L.~R. Rabiner and B.~H. Juang.
\newblock An introduction to hidden {Markov} models.
\newblock {\em IEEE ASSP Magazine}, January:4--16, 1986.

\bibitem{Birney01}
E.~{Birney}.
\newblock Hidden {Markov} models in biological sequence analysis.
\newblock {\em IBM J. Res. Dev.,}, 45(3.4):449--454, 2001.

\bibitem{Eddy04}
S.~Eddy.
\newblock What is a hidden {Markov} model?
\newblock {\em Nature Biotech.}, 22:1315–1316, Oct 2004.

\bibitem{Breto2009}
C.~Bret\'{o}, D.~He, E.~L. Ionides, and A.~A. King.
\newblock Time series analysis via mechanistic models.
\newblock {\em Ann. App. Statistics}, 3(1):319–348, Mar 2009.

\bibitem{Ryden98}
T.~Ryd\'{e}n, T.~Ter\"{a}svirta, and S.~\r{A}sbrink.
\newblock Stylized facts of daily return series and the hidden {Markov} model.
\newblock {\em J. App. Econometrics}, 13:217--244, 1998.

\bibitem{Cove06a}
T.~M. Cover and J.~A. Thomas.
\newblock {\em Elements of Information Theory}.
\newblock Wiley-Interscience, New York, second edition, 2006.

\bibitem{Elton1987}
J.~H. Elton.
\newblock An ergodic theorem for iterated maps.
\newblock {\em Ergod. Th. Dynam. Sys.}, 7:481--488, 1987.

\bibitem{Blac57b}
D.~Blackwell.
\newblock The entropy of functions of finite-state {Markov} chains.
\newblock In {\em Transactions of the first Prague conference on information
  theory, Statistical decision functions, Random processes}, volume~28, pages
  13--20, Prague, Czechoslovakia, 1957. Publishing House of the Czechoslovak
  Academy of Sciences.

\bibitem{Reny59a}
A.~Renyi.
\newblock On the dimension and entropy of probability distributions.
\newblock {\em Acta Math. Hung.}, 10:193, 1959.

\bibitem{Mand82a}
B.~B. Mandelbrot.
\newblock {\em The Fractal Geometry of Nature}.
\newblock W. H. Freeman and Company, San Francisco, California, 1982.

\bibitem{Edga90a}
G.~A. Edgar.
\newblock {\em Measure, Topology, and Fractal Geometry}.
\newblock Springer-Verlag, New York, 1990.

\bibitem{Falc90a}
K.~Falconer.
\newblock {\em Fractal geometry: mathematical foundations and applications}.
\newblock John Wiley, Chichester, 1990.

\bibitem{Pesi97a}
Ya.~B. Pesin.
\newblock {\em Dimension Theory in Dynamical Systems: {Contemporary} Views and
  Applications}.
\newblock University of Chicago Press, 1997.

\bibitem{Shim79a}
I.~Shimada and T.~Nagashima.
\newblock A numerical approach to ergodic problem of dissipative dynamical
  systems.
\newblock {\em Prog. Theo. Phys.}, 61:1605, 1979.

\bibitem{Bene80a}
G.~Benettin, L.~Galgani, A.~Giorgilli, and J.-M. Strelcyn.
\newblock Lyapunov characteristic exponents for smooth dynamical systems and
  for hamiltonian systems; a method for computing all of them.
\newblock {\em Meccanica}, 15:9, 1980.

\bibitem{Kapl79a}
J.~Kaplan and J.~Yorke.
\newblock Chaotic behavior of multidimensional difference equations.
\newblock In {\em Functional Differential Equations and Approximation of Fixed
  Points}, volume 730 of {\em Lecture Notes in Mathematics}, pages 204--227.
  Springer, 1979.

\bibitem{Baran15b}
B.~Barany.
\newblock On the {Ledrappier-Young} formula for self-affine measures.
\newblock {\em Math. Proc. Cambridge Phil. Soc.}, 159(3):405--432, 2015.

\bibitem{Marz17a}
S.~E. Marzen and J.~P. Crutchfield.
\newblock Nearly maximally predictive features and their dimensions.
\newblock {\em Phys. Rev. E}, 95(5):051301(R), 2017.

\bibitem{Vene19a}
A.~Venegas-Li, A.~Jurgens, and J.~P. Crutchfield.
\newblock Measurement-induced randomness and structure in controlled qubit
  processes.
\newblock {\em Physical Review E}, 102(4):040102(R), 2020.

\bibitem{Jurg20a}
A.~Jurgens and J.~P. Crutchfield.
\newblock Functional thermodynamics of {Maxwellian} ratchets: Constructing and
  deconstructing patterns, randomizing and derandomizing behaviors.
\newblock {\em Phys. Rev. Research}, 2(3):033334, 2020.

\end{thebibliography}
\end{document}